\documentclass[preprint]{ptptex} 

\usepackage{wrapft}
\usepackage{amsmath,amssymb,amscd,amsbsy,amsgen,amsopn,amstext,amsxtra}
\usepackage[mathscr]{eucal}

\notypesetlogo                       

\title{Canonical formalism and quantization of a massless spinning bosonic particle 
in four dimensions}

\author{Shinichi \textsc{Deguchi},$^{1,2,}$\footnote{E-mail:  deguchi@phys.cst.nihon-u.ac.jp (SD)} 
Shouma \textsc{Negishi},$^{2}$
Satoshi \textsc{Okano},$^{2,}$\footnote{E-mail:  okano@phys.cst.nihon-u.ac.jp (SO)} 
\\
and 
Takafumi \textsc{Suzuki} $\!{}^{3,}$\footnote{E-mail:  takafumi@gaea.jcn.nihon-u.ac.jp (TS)} 
}

\inst{
$^1$Institute of Quantum Science, College of Science and Technology, \\ 
Nihon University, Chiyoda-ku, Tokyo 101-8308, Japan
\\
$^2$Department of Quantum Science and Technology, \\ 
Graduate School of Science and Technology, \\
Nihon University, Chiyoda-ku, Tokyo 101-8308, Japan
\\
$^3$Junior College Funabashi Campus, \\
Nihon University, Narashinodai, Funabashi, Chiba 274-8501, Japan
}
\abst{%
A twistor model of a free massless spinning particle in 4-dimensional Minkowski space  
is studied in terms of spacetime and spinor variables.  
This model is specified by a simple action,  
referred to here as the gauged Shirafuji action,  
that consists of twistor variables and gauge fields on the 1-dimensional parameter space. 
We consider the canonical formalism of the model by following 
the Dirac formulation for constrained Hamiltonian systems. 
In the subsequent quantization procedure, we obtain a plane-wave solution 
with momentum spinors. 
From this solution and coefficient functions, we construct positive-frequency and negative-frequency 
spinor wave functions defined on complexified Minkowski space. 
It is shown that the Fourier-Laplace transforms of the coefficient functions  
lead to the spinor wave functions expressed as 
the Penrose transforms of the corresponding holomorphic functions on twistor space. 
We also consider the exponential generating function for the spinor wave functions and 
derive a novel representation for each of the spinor wave functions. }

\begin{document} 
\maketitle

\numberwithin{equation}{section}

\section{Introduction} 

Various classical-mechanical models to describe relativistic spinning particles 
have been presented until recently. 
These models can be mainly classified into two types: bosonic models using only 
commutative variables and supersymmetric models using both of commutative and anticommutative variables. 
Rigid body models \cite{Fre,Nak,HG}, relativistic rotator models \cite{Tak1,Tak2}, 
Barut--Zanghi models \cite{BZ,Vaz,KR}, and point-particle models with rigidity \cite{Ply1,Ply2, DerNer} 
are examples of the former type, 
while spinning particle models with world-line supersymmetry \cite{BDZDH,BCL,BM} 
and superparticle models with target-space supersymmetry \cite{Cas1,Cas2,BS} are examples 
of the latter type. 
In this paper, we treat a twistor model of a massless spinning bosonic particle 
in 4-dimensional Minkowski space. This model is an illustrative example of the former type. 
(It is indeed possible to formulate a supersymmetric extension of the twistor model;  
however, we are not concerned with it in this paper.)

A twistor model of a massless spinning bosonic particle 
was first studied by Shirafuji \cite{Shi}. 
His approach is quite natural 
because twistor theory provides elegant and useful descriptions of 
4-dimensional massless systems \cite{PM,PR,HT,Tak}.  
Shirafuji found a simple action integral, referred to here as the {\em Shirafuji action}, 
written in terms of twistor variables. 
The Shirafuji action describes a free massless spinning particle propagating in 4-dimensional Minkowski space   
in a manifestly conformally covariant fashion. After the canonical quantization procedure, 
the canonical commutation relations in twistor quantization \cite{Pen,PM,PR,DN}  
can be reproduced from the corresponding classical Poisson brackets defined from the Shirafuji action. 
(Supersymmetric generalizations of the Shirafuji action have been 
explored in Ref.~\citen{Shi, GumSor, Ban, Ply3, BLS, FIL}, in which supertwistor variables \cite{Ferber} are used to  
describe the dynamics of massless superparticles.)

Recently, the Shirafuji action has been modified in accordance with the gauge principle  
so that it can remain invariant under 
the complexified {\em local} scale transformation of the twistor variables in the action \cite{DEN}. 
This modification is accomplished by gauging the Shirafuji action 
with the aid of gauge fields on the world-line parameter space 
and by adding a one-dimensional Chern-Simons term. 
Because of the complexified local scale invariance, the modified action, which we refer to as 
the {\em gauged Shirafuji action}, is considered to be defined for 
the projective twistor variables rather than the twistor variables.  
This is consistent with the fact that in twistor theory, projective twistors are treated as 
more essential quantities than twistors \cite{PR,HT}.  
The gauged Shirafuju action includes a helicity constraint term due to the modification. 
Hence, it follows that this action describes a free massless spinning particle with a fixed value of helicity.

Although the gauged Shirafuji action written in terms of twistor variables provides an elegant formulation 
for free massless spinning particles, its relations to the ordinary spacetime formulation 
are not sufficiently clear in a sense.  
Also, it seems difficult to incorporate interactions with gauge fields lying in spacetime into 
the gauged Shirafuji action written in terms of twistor variables. 
Considering this situation, in the present paper, we first rewrite the gauged Shirafuji action 
in terms of spacetime and spinor variables. Then we study the canonical Hamiltonian formalism based on 
the gauged Shirafuji action by following the recipe for treating constrained Hamiltonian systems \cite{Dir,HRT,HenT}. 
After a detailed analysis of the constraints in phase space, we perform the canonical quantization of 
our Hamiltonian system and obtain a set of differential equations satisfied 
by a function of spacetime and momentum-spinor variables.   
These equations are solved simultaneously to yield a plane-wave solution with momentum spinors. 
From this solution and coefficient functions, we construct positive-frequency and negative-frequency 
spinor wave functions by taking integrations over the momentum-spinor variables. 
There, convergence of each integral is considered in detail.  As a result,  
it turns out that the spinor wave functions are well-defined 
on their respective tube domains in complexified Minkowski space.  
It is also seen that these functions satisfy generalized Weyl equations.  
Furthermore, we demonstrate that the Fourier-Laplace transforms of the coefficient functions  
lead to the spinor wave functions expressed as 
the Penrose transforms \cite{PM, PR, HT, Tak, Pen} of the corresponding holomorphic functions on twistor space. 
In this way, we can find Penrose transforms via appropriate Fourier-Laplace transforms. 
In addition, we construct the exponential generating function for the spinor wave functions,    
showing that it fulfills the complexification of a fundamental equation 
called the unfolded equation \cite{FedIva,Vas}. 
From the generating function, we derive a novel representation for each of the spinor wave functions, 
which representation is shown to be written in terms of conformally inverted spacetime variables. 
This paper concentrates on investigating free particles; 
interactions with gauge fields in spacetime will be discussed elsewhere.

This paper is organized as follows: 
Section 2 provides a brief review on the gauged Shirafuji action. 
The canonical Hamiltonian formalism based on the gauged Shirafuji action is studied in Sect. 3, and 
the subsequent canonical quantization procedure is performed in Sect. 4. 
In Sect. 5, we construct a positive-frequency spinor wave function and its negative-frequency counterpart  
individually and express them in the form of Penrose transforms. 
In Sect. 6, we consider the exponential generating function for the spinor wave functions 
and derive a novel representation for each of the spinor wave functions. 
Section 7 is devoted to a summary and discussion. 

\newpage

\section{Gauged Shirafuji action}

In this section, we introduce the gauged Shirafuji action and 
rewrite it in terms of spacetime and spinor variables.

Let $Z^{A}=\big(\omega^{\alpha}, \pi_{\dot{\alpha}} \big)$ 
$\big(A=0,1,2,3 \:\! ;\:\! \alpha=0,1;\:\! \dot{\alpha}=\dot{0}, \dot{1} \big)$ be a twistor and 
$\bar{Z}_{A}=\big(\bar{\pi}_{\alpha}, \bar{\omega}{}^{\dot{\alpha}} \big)$ its dual twistor.  
The gauged Shirafuji action is given by 
\begin{align}
S=\int_{\tau_0}^{\tau_1} d\tau 
\bigg[\;\! \frac{i}{2} \lambda  
\big(\bar{Z}_{A} DZ^{A} -Z^{A} \bar{D}\bar{Z}_{A} \big) -2sa\bigg] 
\label{2.1}
\end{align}
with $D:=d/d\tau -ia$ (see Appendix B of Ref. \citen{DEN}). 
Here, $Z^{A}=Z^{A}(\tau)$ and $\bar{Z}_{A}=\bar{Z}_{A}(\tau)$ are understood as complex scalar fields 
on the 1-dimensional parameter space  
$\mathcal{T}:=\{ \:\! \tau\;\! |\, \tau_{0} \leq \tau \leq \tau_{1} \}$.   
Also, $\lambda=\lambda(\tau)$ is a real scalar field on $\mathcal{T}$, 
$a=a(\tau)$ is a real scalar-density field on $\mathcal{T}$, 
and $s$ is a real constant that specifies the helicity of a particle. 
The action $S$ describes a free massless spinning particle propagating 
in 4-dimensional Minkowski space, $\mathbf{M}$, with  
the metric tensor $\eta_{\mu\nu}=\mathrm{diag}(1,-1,-1,-1)$. 
Obviously, $S$ remains invariant under the reparametrization 
$\tau \rightarrow \tau^{\prime} (\tau)$. 
In addition, $S$ remains invariant under the complexified local scale transformation 
\begin{align}
Z^{A} &\rightarrow Z^{\prime A}=\upsilon(\tau) Z^{A} \,, 
\qquad 
\bar{Z}_{A} \rightarrow \bar{Z}^{\prime}_{A}
=\bar{\upsilon}(\tau) \bar{Z}_{A} \,, 
\label{2.2}
\end{align}
supplemented by the transformations  
\begin{align}
\lambda \rightarrow \lambda^{\prime}=|\upsilon(\tau)|^{-2} \lambda \,, 
\qquad
a \rightarrow a^{\prime}=a+\frac{d\theta(\tau)}{d\tau} \,, 
\label{2.3}
\end{align}
where $\upsilon$ is a complex gauge function of $\tau$, and 
$\theta$ is defined by $\theta:={1\over2}i \ln (\bar{\upsilon}/ \upsilon)$. 
Here, note that $a$ behaves as a $U(1)$ gauge field on $\mathcal{T}$     
and that $D$ is treated as its  associated covariant derivative. 
The gauge invariance of the 1-dimensional Chern-Simons term  
$-2s \int_{\tau_0}^{\tau_1} d\tau a$ is ensured by imposing 
an appropriate boundary condition such as $\theta(\tau_{1})=\theta(\tau_{0})$. 
The fact that $S$ remains invariant under the simultaneous transformations (\ref{2.2}) and (\ref{2.3}) 
implies that $S$ is really defined for the proportionality class 
called the {\em projective} twistor, 
$[ Z^{A} ] :=\big\{ c Z^{A} \big|\, c \in \Bbb{C}\setminus\{0\} \big\}$, 
rather than the (nonzero) twistor $Z^{A}$ itself.  
The action $S$ is thus considered to be described with the projective twistor $[ Z^{A} ]$. 
This statement is consistent with the fact that in twistor theory,  
projective twistors are treated as more essential quantities than twistors.  
The projective twistor $[ Z^{A} ]$ is abbreviated as $Z^{A}$ by regarding  
it as a representative element of the set $[ Z^{A} ]$.

Assuming that $\lambda>0$, now we carry out the scaling 
$Z^{A} \rightarrow \lambda^{-1/2} Z^{A}$ and 
$\bar{Z}_{A} \rightarrow \lambda^{-1/2} \bar{Z}_{A}$ to express the action $S$ as 
\begin{align}
S &=\int_{\tau_0}^{\tau_1} d\tau 
\bigg[\;\! \frac{i}{2} 
\big(\bar{Z}_{A} DZ^{A} -Z^{A} \bar{D}\bar{Z}_{A} \big) -2sa\bigg] 
\notag
\\
&=\int_{\tau_0}^{\tau_1} d\tau 
\bigg[\;\! \frac{i}{2} \Big( \bar{Z}_{A} \dot{Z}^{A}-Z^{A} \Dot{\Bar{Z}}_{A} \Big)
+a \big( \bar{Z}_{A} Z^{A} -2s \big) \bigg] \,, 
\label{2.4}
\end{align}
where a dot over a variable denotes its derivative with respect to $\tau$. 
This form of $S$ can also be regarded as Eq. (\ref{2.1}) in a particular gauge $\lambda=1$. 
In this gauge, the pure local scale invariance of $S$ is not observed, while  
the local $U(1)$ gauge invariance of $S$ is still observed in addition to 
the reparametrization invariance.

Here, we would like to make remarks on the field $a$ and the term proportional to it, 
namely $a \big( \bar{Z}_{A} Z^{A} -2s \big)$, included in $S$.   
The possibility of introducing such a term into the (original) Shirafuji action was already  
mentioned in the case $s=0$ by Gumenchuk and Sorokin \cite{GumSor}. 
Plyushchay considered the same term (and its supersymmetric extension) in 
a Hamiltonian formulation of (super)particles\cite{Ply3}. 
Bandos, Lukierski and Sorokin introduced this kind of term into a superparticle model 
to incorporate the (extended) helicity constraint into the action\cite{BLS}.  
However, in these works, 
the fields corresponding to $a$ are merely treated as Lagrangian multipliers, although 
it was pointed out in Ref. \citen{BLS} that the (extended) helicity constraint generates 
a $U(1)$ gauge transformation. 
In Refs. \citen{FIL} and \citen{FedIva}, 
their authors presented models of massless higher-spin (super)particles. 
In these models, generalizations of the term proportional to $a$ are involved to describe (super)particles 
with fixed helicity, 
and the fields corresponding to $a$ are regarded as $U(1)$ gauge fields for 
the local phase transformations of the relevant fields. 
In Ref. \citen{DEN} and in the present paper, unlike the previous approaches, 
$a$, together with $\lambda$, has been introduced 
in the beginning in accordance with the gauge principle so that the action can remain invariant under 
the local transformation (\ref{2.2}). 
The covariant derivative $D$ has been defined accordingly 
and the term proportional to $a$ has been incorporated into $S$ automatically. 
It should be mentioned here that the action (\ref{2.4}) and its local $U(1)$ gauge invariance have also been    
found by Bars and Pic\'{o}n \cite{BP,Bars}.

As seen in the literature on twistor theory \cite{PM,PR,HT}, 
the 2-component spinor $\omega^{\alpha}$ is related with another 2-component spinor 
$\pi_{\dot{\alpha}}$ by 
\begin{align}
\omega^{\alpha}=i z^{\alpha \dot{\alpha}} \pi_{\dot{\alpha}} \,, 
\label{2.5}
\end{align}
where $z^{\alpha \dot{\alpha}}$ are coordinates of a point in complexified 
Minkowski space $\Bbb{C}\mathbf{M}$, being now treated as scalar fields on $\mathcal{T}$. 
The coordinates $z^{\alpha \dot{\alpha}}$ can be decomposed 
as $z^{\alpha \dot{\alpha}}=x^{\alpha \dot{\alpha}}-i y^{\alpha \dot{\alpha}}$, 
where $x^{\alpha \dot{\alpha}}$ and $y^{\alpha \dot{\alpha}}$ 
are elements of Hermitian matrices and hence fulfill the Hermitian conditions 
$\overline{x^{\beta \dot{\alpha}}}=x^{\alpha \dot{\beta}}$ and  
$\overline{y^{\beta \dot{\alpha}}}=y^{\alpha \dot{\beta}}$. 
The matrix elements $x^{\alpha \dot{\alpha}}$ are identified with coordinates of a point in 
Minkowski space $\mathbf{M}$. 
In terms of the spacetime variables $x^{\alpha\dot{\alpha}}$ 
and the spinor variables $\bar{\pi}_{\alpha}$, $\pi_{\dot{\alpha}}$, 
$\psi^{\alpha}:=y^{\alpha\dot{\alpha}} \pi_{\dot{\alpha}}$ and 
$\bar{\psi}^{\dot{\alpha}}:=y^{\alpha\dot{\alpha}} \bar{\pi}_{\alpha}$, 
the action $S$ can be written, 
up to the boundary terms at $\tau_{0}$ and $\tau_{1}$,  as 
\begin{align}
S=\int_{\tau_0}^{\tau_1} d\tau L
\label{2.6}
\end{align}
with the Lagrangian 
\begin{align}
L :=-\dot{x}^{\alpha\dot{\alpha}} \bar{\pi}_{\alpha} \pi_{\dot{\alpha}} 
-i \big( \psi^{\alpha} \Dot{\Bar{\pi}}_{\alpha}
-\bar{\psi}{}^{\dot{\alpha}} \dot{\pi}_{\dot{\alpha}} \big) 
+a \big( \psi^{\alpha} \Bar{\pi}_{\alpha}
+\bar{\psi}{}^{\dot{\alpha}} \pi_{\dot{\alpha}} -2s\big) \,. 
\label{2.7}
\end{align}
This is the gauged Shirafuji action written in terms of spacetime and spinor variables. 
Equation (\ref{2.5}) can be written as 
$\omega^{\alpha}=ix^{\alpha\dot{\alpha}}\pi_{\dot{\alpha}}+\psi^{\alpha}$.\footnote{~Using 
the twistor 
$Y^{A}:=\big(\psi^{\alpha}, \pi_{\dot{\alpha}} \big)$ and its dual twistor 
$\bar{Y}_{A}:=\big(\bar{\pi}_{\alpha}, \bar{\psi}{}^{\dot{\alpha}} \big)$, 
the action $S$ can be expressed in the {\em partially} twistorial form 
\begin{align*}
S &=\int_{\tau_0}^{\tau_1} d\tau 
\bigg[ -\dot{x}^{\alpha\dot{\alpha}} \bar{\pi}_{\alpha} \pi_{\dot{\alpha}} 
+\frac{i}{2} 
\big( \bar{Y}_{A} DY^{A} -Y^{A} \bar{D}\bar{Y}_{A} \big) -2sa\bigg] \,.
\end{align*}
}
The canonical momentum conjugate to $x^{\alpha\dot{\alpha}}$ is found to be  
$P^{(x)}_{\alpha\dot{\alpha}}:=\partial{L}/\partial \dot{x}^{\alpha\dot{\alpha}}
=-\bar{\pi}_{\alpha} \pi_{\dot{\alpha}}\:\!$, 
which ensures that $\bar{\pi}_{\alpha}$ and $\pi_{\dot{\alpha}}$  
are, as we say, momentum spinors. It follows from 
$\bar{\pi}_{\alpha} \bar{\pi}^{\alpha}=\pi_{\dot{\alpha}} \pi^{\dot{\alpha}}=0$ 
that $P_{\alpha\dot{\alpha}} P^{\alpha\dot{\alpha}}=0$. 
This shows that the action $S$ describes a massless particle.

\section{Canonical formalism}

In this section, we study the canonical Hamiltonian formalism of the model governed by the action $S$. 
We treat the variables 
$\left( x^{\alpha\dot{\alpha}}, \bar{\pi}_{\alpha}, \pi_{\dot{\alpha}}, 
\psi^{\alpha}, \bar{\psi}^{\dot{\alpha}}, a \right)$ contained in the Lagrangian $L$ as canonical coordinates. 
Their corresponding conjugate momenta are defined by 
\begin{subequations}
\label{3.1}
\begin{align}
P_{\alpha \dot{\alpha}}^{(x)} 
&:= \frac{\partial L}{\partial \dot{x}^{\alpha \dot{\alpha}}}
= -\bar{\pi}_\alpha \pi_{\dot{\alpha}} \,, 
\label{3.1a}
\\
P_{(\bar{\pi})}^{\:\!\alpha}
&:= \frac{\partial L}{\partial \dot{\bar{\pi}}_\alpha}
= -i \psi^{\alpha} , 
\label{3.1b}
\\
P_{(\pi)}^{\:\!\dot{\alpha}}
&:= \frac{\partial L}{\partial \dot{\pi}_{\dot{\alpha}}}
= i \bar{\psi}{}^{\dot{\alpha}} , 
\label{3.1c}
\\
P^{(\psi)}_{\alpha}
&:= \frac{\partial L}{\partial \dot{\psi}^\alpha}
= 0 \,, 
\label{3.1d}
\\
P^{(\bar{\psi})}_{\dot{\alpha}}
&:= \frac{\partial L}{\partial \dot{\bar{\psi}}^{\dot{\alpha}}}
= 0 \,, 
\label{3.1e}
\\
P^{(a)}
&:= \frac{\partial L}{\partial \dot{a}}
= 0 \,.
\label{3.1f}
\end{align}
\end{subequations}
The canonical Hamiltonian is found from Eqs. (\ref{2.7}) and (\ref{3.1}) to be 
\begin{align}
H_{\rm{C}} &:=\dot{x}^{\alpha \dot{\alpha}} P_{\alpha \dot{\alpha}}^{(x)}
+\dot{\bar{\pi}}_{\alpha} P_{(\bar{\pi})}^{\:\!\alpha} +\dot{\pi}_{\dot{\alpha}} P_{(\pi)}^{\:\!\dot{\alpha}}
+\dot{\psi}^{\alpha} P^{(\psi)}_{\alpha} +\dot{\bar{\psi}}{}^{\dot{\alpha}} P^{(\bar{\psi})}_{\dot{\alpha}}
+\dot{a} P^{(a)} - L 
\notag \\
&\;=-a \big( \psi^{\alpha} \Bar{\pi}_{\alpha}
+\bar{\psi}{}^{\dot{\alpha}} \pi_{\dot{\alpha}} -2s\big) \,. 
\label{3.2}
\end{align}
The non-vanishing Poisson brackets between the canonical variables are given by 
\begin{alignat}{5}
\left\{ x^{\alpha \dot{\alpha}} , P_{\beta \dot{\beta}}^{(x)} \right\} 
&= \delta_{\beta}^{\alpha} \delta_{\dot{\beta}}^{\dot{\alpha}} \,, 
&\quad 
\left\{ \bar{\pi}_\alpha , P^{\:\!\beta}_{(\bar{\pi})} \right\} &= \delta_{\alpha}^{\beta} \,, 
&\quad
\left\{ \pi_{\dot{\alpha}} , P^{\:\!\dot{\beta}}_{({\pi})} \right\} &= \delta_{\dot{\alpha}}^{\dot{\beta}} \,, 
\notag
\\
\left\{ \psi^\alpha , P_\beta^{(\psi)} \right\} &= \delta^{\alpha}_{\beta} \,,
&\quad 
\left\{ \bar{\psi}{}^{\dot{\alpha}} , P_{\dot{\beta}}^{(\bar{\psi})} \right\} &= \delta^{\dot{\alpha}}_{\dot{\beta}} \,,
&\quad
\left\{ a \:\!, P^{(a)} \right\} &= 1 \,. 
\label{3.3}
\end{alignat}
The Poisson bracket between two arbitrary analytic functions of the canonical variables  
can be calculated using the fundamental Poisson brackets in Eq. (\ref{3.3}).

Equations (\ref{3.1a})--(\ref{3.1f}) are read, respectively, as the primary constraints 
\begin{subequations}
\label{3.4}
\begin{align}
\phi_{\alpha \dot{\alpha}}^{(x)} 
&:= P_{\alpha \dot{\alpha}}^{(x)} + \bar{\pi}_\alpha \pi_{\dot{\alpha}} \approx 0 \,, 
\label{3.4a} 
\\
\phi^{\alpha}_{(\bar{\pi})} 
&:= P^{\:\!\alpha}_{(\bar{\pi})} + i \psi^{\alpha} \approx 0 \,,
\label{3.4b} 
\\
\phi^{\dot{\alpha}}_{(\pi)} 
&:= P^{\:\!\dot{\alpha}}_{(\pi)} - i \bar{\psi}{}^{\dot{\alpha}} \approx 0 \,, 
\label{3.4c}
\\
\phi^{(\psi)}_{\alpha} 
&:= P^{(\psi)}_{\alpha} \approx 0 \,, 
\label{3.4d}
\\
\phi^{(\bar{\psi})}_{\dot{\alpha}} 
&:= P^{(\bar{\psi})}_{\dot{\alpha}} \approx 0 \,, 
\label{3.4e}
\\
\phi^{(a)}	
&:= P^{(a)} \approx 0 \,.
\label{3.4f}
\end{align}
\end{subequations}
where the symbol g$\approx$h denotes the weak equality. Now, we apply the Dirac formulation for
constrained Hamiltonian systems \cite{Dir, HRT, HenT} to the present model. 
To this end, we first calculate the Poisson brackets between the constraint functions 
$\phi_{\alpha \dot{\alpha}}^{(x)}$, $\phi^{\alpha}_{(\bar{\pi})}$, $\phi^{\dot{\alpha}}_{(\pi)}$, 
$\phi^{(\psi)}_{\alpha}$, $\phi^{(\bar{\psi})}_{\dot{\alpha}}$ and $\phi^{(a)}$, 
obtaining the following non-vanishing Poisson brackets:  
\begin{alignat}{3} 
\left\{ \phi_{\alpha \dot{\alpha}}^{(x)} , \phi^{\beta}_{(\bar{\pi})} \right\}
&= \delta_{\alpha}^{\beta} \pi_{\dot{\alpha}} \,, 
& \quad \;\;
\left\{ \phi_{\alpha \dot{\alpha}}^{(x)} , \phi^{\dot{\beta}}_{({\pi})} \right\} 
& = \bar{\pi}_{\alpha} \delta_{\dot{\alpha}}^{\dot{\beta}} \,, 
\notag
\\
\left\{ \phi_{(\bar{\pi})}^{\alpha} , \phi_{\beta}^{(\psi)} \right\}
& = i  \delta^{\alpha}_{\beta} \,,
& \quad \;\;
\left\{ \phi_{(\pi)}^{\dot{\alpha}} , \phi_{\dot{\beta}}^{(\bar{\psi})} \right\} 
& = -i \delta^{\dot{\alpha}}_{\dot{\beta}} \,.
\label{3.5}
\end{alignat}
We can also obtain
\begin{alignat}{5}
\left\{ \phi_{\alpha \dot{\alpha}}^{(x)} , H_{\rm{C}} \right\}
& = 0 \,, 
& \quad 
\left\{ \phi^{\alpha}_{(\bar{\pi})} , H_{\rm{C}} \right\} 
& = a \psi^\alpha \,,
& \quad 
\left\{ \phi^{\dot{\alpha}}_{({\pi})} , H_{\rm{C}} \right\}
& = a \bar{\psi}^{\dot{\alpha}} \,,
\notag
\\
\left\{ \phi_{\alpha}^{(\psi)} , H_{\rm{C}} \right\}
& = a \bar{\pi}_\alpha \,,
& \quad
\left\{ \phi_{\dot{\alpha}}^{(\bar{\psi})} , H_{\rm{C}} \right\} 
& = a \pi_{\dot{\alpha}} \,,
& \quad
\left\{ \phi^{(a)} , H_{\rm{C}} \right\} 
& = \psi^\alpha \bar{\pi}_\alpha + \bar{\psi}{}^{\dot{\alpha}}\pi_{\dot{\alpha}} -2s \,.
\label{3.6}
\end{alignat}
Introducing the Lagrange multipliers 
$u^{\alpha \dot{\alpha}}_{(x)}$, $u_{\alpha}^{(\bar{\pi})}$, $u_{\dot{\alpha}}^{(\pi)}$, 
$u_{(\psi)}^{\alpha}$, $u_{(\bar{\psi})}^{\dot{\alpha}}$ and $u_{(a)}$, 
we define the total Hamiltonian 
\begin{align}
H_{\rm T}:=H_{\rm C}
+u^{\alpha \dot{\alpha}}_{(x)} \phi_{\alpha \dot{\alpha}}^{(x)}
+u_{\alpha}^{(\bar{\pi})} \phi^{\alpha}_{(\bar{\pi})}
+u_{\dot{\alpha}}^{(\pi)} \phi^{\dot{\alpha}}_{(\pi)}
+u_{(\psi)}^{\alpha} \phi^{(\psi)}_{\alpha} 
+u_{(\bar{\psi})}^{\dot{\alpha}} \phi^{(\bar{\psi})}_{\dot{\alpha}}
+u_{(a)} \phi^{(a)} .
\label{3.7}
\end{align}
With this Hamiltonian,  the canonical equation for a function $f$ of the canonical variables 
is given by 
\begin{align}
\dot{f}=\{ f, H_{\rm T} \} \,.
\label{3.8}
\end{align}
The primary constraints (\ref{3.4a})--(\ref{3.4f}) must be preserved in time, because they are valid at any time. 
The time evolutions of the constraint functions can be evaluated using Eqs. (\ref{3.4})--(\ref{3.8}), 
and as a result, we have the consistency conditions  
\begin{subequations}
\label{3.9}
\begin{align}
\dot{\phi}_{\alpha \dot{\alpha}}^{(x)}
 &= \left\{ \phi_{\alpha \dot{\alpha}}^{(x)} , H_{\rm{T}} \right\}
\approx u_{\alpha}^{(\bar{\pi})} \pi_{\dot{\alpha}} + u_{\dot{\alpha}}^{({\pi})} \bar{\pi}_\alpha
\approx 0 \,,
\label{3.9a} 
\\
\dot{\phi}_{(\bar{\pi})}^{\alpha}
&= \left\{ \phi_{(\bar{\pi})}^{\alpha} , H_{\rm{T}} \right\}
\approx a \psi^\alpha -u^{\alpha \dot{\alpha}}_{(x)} \pi_{\dot{\alpha}} + i u_{(\psi)}^{\alpha}
\approx 0 \,,
\label{3.9b} 
\\
\dot{\phi}_{(\pi)}^{\dot{\alpha}}
&= \left\{ \phi_{(\pi)}^{\dot{\alpha}} , H_{\rm{T}} \right\}
\approx a \bar{\psi}^{\dot{\alpha}} -u^{\alpha \dot{\alpha}}_{(x)} \bar{\pi}_{\alpha} - i u_{(\bar{\psi})}^{\dot{\alpha}} 
\approx 0 \,, 
\label{3.9c} 
\\
\dot{\phi}_{\alpha}^{(\psi)}
&= \left\{ \phi_{\alpha}^{(\psi)} , H_{\rm{T}} \right\}
\approx a \bar{\pi}_{\alpha} - i u_{\alpha}^{(\bar{\pi})}  
\approx 0 \,,
\label{3.9d} 
\\
\dot{\phi}_{\dot{\alpha}}^{(\bar{\psi})}
&= \left\{ \phi_{\dot{\alpha}}^{(\bar{\psi})} , H_{\rm{T}} \right\}
\approx a \pi_{\dot{\alpha}} + i u_{\dot{\alpha}}^{(\pi)} 
\approx 0 \,,
\label{3.9e} 
\\
\dot{\phi}^{(a)}
&= \left\{ \phi^{(a)} , H_{\rm{T}} \right\} 
= \psi^{\alpha} \bar{\pi}_{\alpha} + \bar{\psi}{}^{\dot{\alpha}} \pi_{\dot{\alpha}} - 2s 
\approx 0 \,.
\label{3.9f}
\end{align}
\end{subequations}
Equations (\ref{3.9d}) and (\ref{3.9e}) determine 
$u_{\alpha}^{(\bar{\pi})}$ and $u_{\dot{\alpha}}^{(\pi)}$, respectively, as   
$u_{\alpha}^{(\bar{\pi})} \approx -ia \bar{\pi}_{\alpha}$ and 
$u_{\dot{\alpha}}^{(\pi)} \approx ia \pi_{\dot{\alpha}}$.  
Substituting these into Eq. (\ref{3.9a}), we see that $\dot{\phi}_{\alpha \dot{\alpha}}^{(x)} \approx 0$ 
is identically satisfied; hence, Eq. (\ref{3.9a}) gives no new constraints. 
The Lagrange multipliers $u_{(\psi)}^{\alpha}$ and $u_{(\bar{\psi})}^{\dot{\alpha}}$ are determined 
from Eqs. (\ref{3.9b}) and (\ref{3.9c}), respectively, if $u^{\alpha \dot{\alpha}}_{(x)}$ is 
fixed to a specific function of the canonical variables. 
Consequently, it turns out that 
$u^{\alpha \dot{\alpha}}_{(x)}$ and $u_{(a)}$ still remain undetermined.  
This implies that Eqs. (\ref{3.4a}) and (\ref{3.4f}) are first-class primary constraints. 
Equation (\ref{3.9f}) gives rise to the secondary constraint 
\begin{align}
\chi^{(a)}:=\psi^{\alpha} \bar{\pi}_{\alpha} + \bar{\psi}{}^{\dot{\alpha}} \pi_{\dot{\alpha}} - 2s 
\approx 0 \,.
\label{3.10}
\end{align}
The non-vanishing Poisson brackets between $\chi^{(a)}$ and the primary constraint functions 
are found to be 
\begin{alignat}{3} 
\left\{ \chi^{(a)},  \phi^{\alpha}_{(\bar{\pi})} \right\}
&= \psi^{\alpha} \,, 
& \quad \;\;
\left\{ \chi^{(a)} , \phi^{\dot{\alpha}}_{({\pi})} \right\} 
& =\bar{\psi}{}^{\dot{\alpha}} \,, 
\notag
\\
\left\{ \chi^{(a)} , \phi_{\alpha}^{(\psi)} \right\}
& = \bar{\pi}_{\alpha}  \,,
& \quad \;\;
\left\{  \chi^{(a)} , \phi_{\dot{\alpha}}^{(\bar{\psi})} \right\} 
& = \pi_{\dot{\alpha}} \,.
\label{3.11}
\end{alignat}
Then the time evolution of $\chi^{(a)}$ is evaluated as 
\begin{align}
\dot{\chi}^{(a)}
 &= \left\{ \chi^{(a)} , H_{\rm{T}} \right\}
\approx u_{\alpha}^{(\bar{\pi})} \psi^{\alpha} 
+ u_{\dot{\alpha}}^{({\pi})} \bar{\psi}{}^{\dot{\alpha}} 
+ u_{(\psi)}^{\alpha} \bar{\pi}_{\alpha}
+ u_{(\bar{\psi})}^{\dot{\alpha}} \pi_{\dot{\alpha}} \,. 
\label{3.12} 
\end{align}
By using Eqs. (\ref{3.9b})--(\ref{3.9e}), we can eliminate the Lagrange multipliers in Eq. (\ref{3.12}) 
and obtain $\dot{\chi}^{(a)}\approx 0$. 
Because this condition is identically satisfied, no new constraints are derived anymore.  
Thus the procedure for deriving constraints is completed.

We have obtained all the non-vanishing Poisson brackets between 
the constraint functions, as in Eqs. (\ref{3.5}) and (\ref{3.11}). 
However, it is difficult to classify the constraints into first and second classes on the basis of 
Eqs. (\ref{3.5}) and (\ref{3.11}) together with the vanishing Poisson brackets 
between the constraint functions. 
To find simpler forms of the relevant Poisson brackets, now we define  
\begin{align}
\tilde{\phi}_{\alpha \dot{\alpha}}^{(x)}
&:= \phi_{\alpha \dot{\alpha}}^{(x)} - i \phi_\alpha^{(\psi)} \pi_{\dot{\alpha}} 
+ i \bar{\pi}_\alpha \phi_{\dot{\alpha}}^{(\bar{\psi})}  ,
\label{3.13} 
\\
\tilde{\chi}^{(a)}
&:= \chi^{(a)} -i \psi^\alpha \phi_\alpha^{(\psi)} + i \bar{\psi}^{\dot{\alpha}} \phi_{\dot{\alpha}}^{(\bar{\psi})}
+ i \bar{\pi}_\alpha \phi_{(\bar{\pi})}^\alpha - i \pi_{\dot{\alpha}} \phi_{(\pi)}^{\dot{\alpha}}  \,. 
\label{3.14}
\end{align}
It is immediately seen that the set of all constraints 
$\left( \phi_{\alpha \dot{\alpha}}^{(x)}, \phi^{\alpha}_{(\bar{\pi})}, \phi^{\dot{\alpha}}_{(\pi)}, 
\phi^{(\psi)}_{\alpha}, \phi^{(\bar{\psi})}_{\dot{\alpha}}, \phi^{(a)}, \chi^{(a)} \right) \approx 0$ 
is equivalent to the new set of constraints 
$\left( \tilde{\phi}_{\alpha \dot{\alpha}}^{(x)}, \phi^{\alpha}_{(\bar{\pi})}, 
\phi^{\dot{\alpha}}_{(\pi)}, \phi^{(\psi)}_{\alpha}, \phi^{(\bar{\psi})}_{\dot{\alpha}}, \phi^{(a)}, \tilde{\chi}^{(a)} \right) \approx 0$. 
We can show that except for 
\begin{align}
\left\{ \phi_{(\bar{\pi})}^{\alpha} , \phi_{\beta}^{(\psi)} \right\}= i  \delta^{\alpha}_{\beta} \,,
\quad \;\;
\left\{ \phi_{(\pi)}^{\dot{\alpha}} , \phi_{\dot{\beta}}^{(\bar{\psi})} \right\} 
= -i \delta^{\dot{\alpha}}_{\dot{\beta}} \,, 
\label{3.15}
\end{align}
all the other Poisson brackets between the constraint functions in the new set vanish. 
In this way, the relevant Poisson brackets are simplified in terms of  
$\tilde{\phi}_{\alpha \dot{\alpha}}^{(x)}$ and $\tilde{\chi}^{(a)}$.  
The Poisson brackets between the constraint functions are summarized in a matrix form as 
\begin{align}
\bordermatrix{
								&	\; \tilde{\phi}_{\beta \dot{\beta}}^{(x)}	&	\phi_{(\bar{\pi})}^\beta	&	\phi_{(\pi)}^{\dot{\beta}}	
								&	\phi_\beta^{(\psi)}	&	\phi_{\dot{\beta}}^{(\bar{\psi})}	&	\phi^{(a)}	&  \tilde{\chi}^{(a)} 	\:\! \cr
  \tilde{\phi}_{\alpha \dot{\alpha}}^{(x)}	&	0	&	0	&	0	&	0	&	0	&	0	&	\!0	\cr
  \phi_{(\bar{\pi})}^\alpha				&	0	&	0	&	0	
								&	i \delta^\alpha_\beta	&	0	&	0	&	\!0	\cr
  \phi_{(\pi)}^{\dot{\alpha}}				&	0	&	0	&	0	
								&	0	&	- i \delta^{\dot{\alpha}}_{\dot{\beta}}	&		0		&		\!0	\cr
  \phi_\alpha^{(\psi)}					&	0	&	- i \delta_\alpha^\beta	&	0
								&	0	&	0	&	0	&	\!0	\cr
  \phi_{\dot{\alpha}}^{(\bar{\psi})}		&	0	&	0	&	i \delta_{\dot{\alpha}}^{\dot{\beta}}
								&	0	&	0	&	0	&	\!0	\cr
  \phi^{(a)}						&	0	&	0	&	0	&	0	&	0	&	0	&	\!0	\cr
  \tilde{\chi}^{(a)}					&	0	&	0	&	0	&	0	&	0	&	0	&	\!0	
 } \, . 
\label{3.16}
\end{align}
We can read from this matrix that $\tilde{\phi}_{\alpha \dot{\alpha}}^{(x)} \approx 0$, 
$\phi^{(a)} \approx 0$, and $\tilde{\chi}^{(a)} \approx 0$ are first-class constraints, 
while $\phi_{(\bar{\pi})}^\alpha \approx 0$, $ \phi_{(\pi)}^{\dot{\alpha}}	 \approx 0$, 
$\phi_\alpha^{(\psi)} \approx 0$, and $\phi_{\dot{\alpha}}^{(\bar{\psi})} \approx 0$ are 
second-class constraints.

Following Dirac's approach to second-class constraints, 
we define the Dirac bracket with the aid of the largest invertible submatrix of the matrix (\ref{3.16}). 
For arbitrary functions $f$ and $g$ of the canonical variables, the Dirac bracket is defined by 
\begin{align}
\left\{ f , g \right\}_{\rm D} := \left\{ f , g \right\} 
  &- i \left\{ f , \phi_{(\bar{\pi})}^\alpha \right\} \! \left\{ \phi_\alpha^{(\psi)} , g \right\} 
  + i \left\{ f , \phi^{\dot{\alpha}}_{(\pi)} \right\} \! \left\{ \phi_{\dot{\alpha}}^{(\bar{\psi})} , g \right\}
\notag 
\\
  &+ i \left\{ f , \phi_\alpha^{(\psi)} \right\} \! \left\{ \phi_{(\bar{\pi})}^\alpha , g \right\} 
  - i \left\{ f , \phi_{\dot{\alpha}}^{(\bar{\psi})} \right\} \!\left\{ \phi^{\dot{\alpha}}_{(\pi)} , g \right\} . 
\label{3.17} 
\end{align}
Because the Dirac bracket between $f$ and each of the functions  
$\phi_{(\bar{\pi})}^\alpha$, $ \phi_{(\pi)}^{\dot{\alpha}}$, 
$\phi_\alpha^{(\psi)}$ and $\phi_{\dot{\alpha}}^{(\bar{\psi})}$ vanishes identically, 
the second-class constraints 
can be set strongly equal to zero and may be expressed as  
$\phi_{(\bar{\pi})}^\alpha=0$, $ \phi_{(\pi)}^{\dot{\alpha}}=0$, 
$\phi_\alpha^{(\psi)}=0$, and $\phi_{\dot{\alpha}}^{(\bar{\psi})}=0$,    
as long as the Dirac bracket $\left\{ f , g \right\}_{\rm D}$ is adopted. 
Accordingly, $\psi^{\alpha}$ and $\bar{\psi}{}^{\dot{\alpha}}$ can be identified with the conjugate 
momenta of $\bar{\pi}_{\alpha}$ and $\pi_{\dot{\alpha}}$, respectively 
(up to multiplicative constants). 
Hereafter, with the Dirac bracket $\left\{ f , g \right\}_{\rm D}$, we treat  
$\left( x^{\alpha\dot{\alpha}}, \bar{\pi}_{\alpha}, \pi_{\dot{\alpha}}, a \right)$ as canonical coordinates 
and treat $\Big( P_{\alpha \dot{\alpha}}^{(x)}, \psi^{\alpha}, \bar{\psi}{}^{\dot{\alpha}}, P^{(a)} \Big)$ 
as their corresponding conjugate momenta. 
The non-vanishing Dirac brackets between these canonical variables are found  
from Eqs. (\ref{3.17}) and (\ref{3.3}) to be  
\begin{alignat}{3}
\left\{ x^{\alpha \dot{\alpha}} , P_{\beta \dot{\beta}}^{(x)} \right\}_{\rm D} \!
&= \delta_{\beta}^{\alpha} \delta_{\dot{\beta}}^{\dot{\alpha}} \,, 
&\quad \;\; 
\left\{ a \:\!, P^{(a)} \right\}_{\rm D} \! &= 1 \,,
\notag
\\
\left\{ \bar{\pi}_\alpha , \psi^{\beta} \right\}_{\rm D} \! &= i \delta_{\alpha}^{\beta} \,, 
&\quad \;\;
\left\{ \pi_{\dot{\alpha}} , \bar{\psi}{}^{\dot{\beta}} \right\}_{\rm D} \! 
&= -i \delta_{\dot{\alpha}}^{\dot{\beta}} \,. 
\label{3.18}
\end{alignat}
Because the second-class constraints are now strong equations, 
Eqs. (\ref{3.13}) and (\ref{3.14}) reduce to 
$\tilde{\phi}_{\alpha \dot{\alpha}}^{(x)}= \phi_{\alpha \dot{\alpha}}^{(x)}$ and 
$\tilde{\chi}^{(a)}= \chi^{(a)}$, respectively. For this reason, it follows that the first-class constraints 
that we should take into account are eventually 
\begin{align}
\phi_{\alpha \dot{\alpha}}^{(x)} \approx 0 \,, \;\quad 
\phi^{(a)} \approx 0 \,,  \;\quad 
\chi^{(a)} \approx 0 \,.
\label{3.19}
\end{align}

\section{Canonical quantization}

In this section, we perform canonical quantization of the Hamiltonian system examined in Sect. 3. 
To this end, in accordance with Dirac's method of quantization, we introduce the operators 
$\hat{f}$ and $\hat{g}$ corresponding to the functions $f$ and $g$, respectively, 
and set the commutation relation 
\begin{align}
\left[\:\! \hat{f}, \hat{g} \:\! \right] \!=i \;\! \widehat {\left\{ f , g \right\} }_{\rm D} 
\label{4.1}
\end{align}
in units such that $\hbar=1$.  
Here, $\widehat {\left\{ f , g \right\} }_{\rm D}$ denotes the operator corresponding to 
the Dirac bracket $\left\{ f , g \right\}_{\rm D}$. 
From Eqs. (\ref{3.18}) and (\ref{4.1}), we have the canonical 
commutation relations 
\begin{alignat}{3}
\left[ \:\! \hat{x}^{\alpha \dot{\alpha}} , \hat{P}_{\beta \dot{\beta}}^{(x)} \right] \!
&= i \delta_{\beta}^{\alpha} \delta_{\dot{\beta}}^{\dot{\alpha}} \,, 
&\quad \;\; 
\left[ \:\! \hat{a} \:\!, \hat{P}^{(a)} \right] \! &= i \,,
\notag
\\
\left[ \:\! \Hat{\Bar{\pi}}_\alpha , \hat{\psi}{}^{\beta} \:\! \right] \! &= -\delta_{\alpha}^{\beta} \,, 
&\quad \;\;
\left[ \:\! \hat{\pi}_{\dot{\alpha}} , \Hat{\Bar{\psi}}{}^{\dot{\beta}} \:\! \right] \! 
&= \delta_{\dot{\alpha}}^{\dot{\beta}} \,. 
\label{4.2}
\end{alignat}
The other canonical commutation relations vanish.

In the quantization procedure, the first-class constraints in Eq. (\ref{3.19}) lead to 
the physical state conditions 
\begin{subequations}
\label{4.3}
\begin{align}
\hat{\phi}_{\alpha \dot{\alpha}}^{(x)} |\varPhi \rangle
&= \Big( \hat{P}_{\alpha \dot{\alpha}}^{(x)} + \Hat{\Bar{\pi}}_\alpha \hat{\pi}_{\dot{\alpha}} \Big) 
|\varPhi \rangle =0 \,, 
\label{4.3a} 
\\
\hat{\phi}^{(a)}	|\varPhi \rangle 
&= \hat{P}^{(a)} |\varPhi \rangle =0 \,, 
\label{4.3b}
\\
\hat{\chi}^{(a)} |\varPhi \rangle 
&=\Big( \Hat{\Bar{\pi}}_{\alpha} \hat{\psi}^{\alpha} 
+\hat{\pi}_{\dot{\alpha}} \Hat{\Bar{\psi}}{}^{\dot{\alpha}}- 2s \Big) |\varPhi \rangle=0 \,, 
\label{4.3c}
\end{align}
\end{subequations}
where $|\varPhi \rangle$ denotes a physical state. 
In defining the operators $\hat{\phi}_{\alpha \dot{\alpha}}^{(x)}$ and $\hat{\chi}^{(a)}$, we have obeyed   
the Weyl ordering rule. Then we have used the relevant canonical commutation relations to simplify 
the Weyl ordered operators.

Now we introduce the bra-vector 
\begin{align}
\langle x, a, \bar{\pi}, \pi |
:=\langle 0 | \exp \!\left( ix^{\alpha\dot{\alpha}} \hat{P}_{\alpha \dot{\alpha}}^{(x)}
+ia \hat{P}^{(a)} 
+\bar{\pi}_{\alpha} \hat{\psi}{}^{\alpha} -\pi_{\dot{\alpha}} \Hat{\Bar{\psi}}{}^{\dot{\alpha}} \right) 
\label{4.4}
\end{align}
with the reference bra-vector $\langle 0 |$ specified by 
\begin{align}
\langle 0 |\:\!\hat{x}{}^{\alpha \dot{\alpha}}=\langle 0 |\hat{a} 
=\langle 0 | \Hat{\Bar{\pi}}_{\alpha} =\langle 0 | \hat{\pi}_{\dot{\alpha}}=0 \,.
\label{4.5}
\end{align}
Using the commutation relations in Eq. (\ref{4.2}), we can show that 
\begin{alignat}{3}
\langle x, a, \bar{\pi}, \pi | \:\!\hat{x}{}^{\alpha \dot{\alpha}}
&=x^{\alpha\dot{\alpha}} \langle x, a, \bar{\pi}, \pi | \,, 
&\quad \;\; 
\langle x, a, \bar{\pi}, \pi | \hat{a} 
&=a \langle x, a, \bar{\pi}, \pi | \,,
\notag
\\
\langle x, a, \bar{\pi}, \pi | \Hat{\Bar{\pi}}_{\alpha} 
&=\bar{\pi}_{\alpha} \langle x, a, \bar{\pi}, \pi | \,, 
&\quad \;\;\;
\langle x, a, \bar{\pi}, \pi | \hat{\pi}_{\dot{\alpha}}
&=\pi_{\dot{\alpha}} \langle x, a, \bar{\pi}, \pi | \,. 
\label{4.6}
\end{alignat}
Also, it is easy to see that 
\begin{alignat}{3}
\langle x, a, \bar{\pi}, \pi | \hat{P}_{\alpha \dot{\alpha}}^{(x)}
&=-i\frac{\partial}{\partial x^{\alpha\dot{\alpha}}} \langle x, a, \bar{\pi}, \pi | \,, 
&\quad \;\;\;  
\langle x, a, \bar{\pi}, \pi | \hat{P}^{(a)}
&=-i\frac{\partial}{\partial a} \langle x, a, \bar{\pi}, \pi | \,,
\notag
\\
\langle x, a, \bar{\pi}, \pi | \hat{\psi}{}^{\alpha} 
&=\frac{\partial}{\partial \bar{\pi}_{\alpha}} \langle x, a, \bar{\pi}, \pi | \,, 
&\quad \;\;\;
\langle x, a, \bar{\pi}, \pi | \Hat{\Bar{\psi}}{}^{\dot{\alpha}}
&=-\frac{\partial}{\partial \pi_{\dot{\alpha}}} \langle x, a, \bar{\pi}, \pi | \,. 
\label{4.7}
\end{alignat}
Multiplying each of Eqs. (\ref{4.3a}), (\ref{4.3b}) and (\ref{4.3c}) by $\langle x, a, \bar{\pi}, \pi|$ on the left  
and using Eqs. (\ref{4.6}) and (\ref{4.7}), we have 
\begin{subequations}
\label{4.8}
\begin{align}
\left( -i\frac{\partial}{\partial x^{\alpha\dot{\alpha}}}
+\bar{\pi}_{\alpha} \pi_{\dot{\alpha}} \right) 
\!\varPhi(x, a, \bar{\pi}, \pi) &=0 \,, 
\label{4.8a}
\\
-i\frac{\partial}{\partial a} \varPhi(x, a, \bar{\pi}, \pi) &=0 \,, 
\label{4.8b}
\\
\left( \bar{\pi}_{\alpha} \frac{\partial}{\partial \bar{\pi}_{\alpha}}
-\pi_{\dot{\alpha}} \frac{\partial}{\partial \pi_{\dot{\alpha}}} -2s \right) 
\!\varPhi(x, a, \bar{\pi}, \pi) &=0 \,, 
\label{4.8c}
\end{align}
\end{subequations}
with 
$\varPhi(x, a, \bar{\pi}, \pi):=\langle x, a, \bar{\pi}, \pi |\varPhi \rangle$. 
Equation (\ref{4.8b}) implies that $\varPhi$ is independent of $a$. 
Equations (\ref{4.8a}) and (\ref{4.8c}) can be solved simultaneously for any arbitrary real constant $s$. 
However, if the solution is required to be a Lorentz spinor consisting only of 
$\bar{\pi}_{\alpha}$, $\pi_{\dot{\alpha}}$ and $x^{\alpha\dot{\alpha}}$, 
it is restricted to 
\begin{align}
\varPhi_{\alpha_1 \ldots \alpha_m \dot{\alpha}_1 \ldots \dot{\alpha}_n} 
(x, \bar{\pi}, \pi) =
\bar{\pi}_{\alpha_1} \cdots \bar{\pi}_{\alpha_m} 
\pi_{\dot{\alpha}_1} \cdots \pi_{\dot{\alpha}_n} 
\! \exp\!\left({-ix^{\beta\dot{\beta}} \bar{\pi}_{\beta} \pi_{\dot{\beta}}} \right) , 
\label{4.9}
\end{align}
and accordingly $s$ is determined to be 
\begin{align}
s=\frac{1}{2}(m-n) \,, \quad\; m,n=0,1,2,\ldots. 
\label{4.10}
\end{align}
In this way, the allowed values of the helicity $s$ turn out to be either integer or half-integer values. 
Since the coordinate time is given by $x^{0}=(x^{0\dot{0}}+x^{1\dot{1}})/\sqrt{2}$, 
we see that Eq. (\ref{4.9}) describes a plane-wave of the {\em positive} frequency 
$(|\pi_{\dot{0}}|^{2}+|\pi_{\dot{1}}|^{2})/\sqrt{2}$. 
A negative-frequency plane-wave function can be obtained by taking the complex conjugate of Eq. (\ref{4.9}).

\section{Spinor wave functions and Penrose transforms}

In this section, we construct positive-frequency and negative-frequency 
spinor wave functions from the plane-wave solution (\ref{4.9}), 
considering a regularization method to have well-defined spinor wave functions. 
We also find Penrose transforms via appropriate Fourier-Laplace transforms.

\subsection{Positive-frequency wave function} 

Let $\tilde{f}^{+}(\bar{\pi}, \pi)$ be a complex function that behaves asymptotically as  
$\tilde{f}_{1}^{+}(\bar{\pi}_{1}, \pi_{\dot{1}}) (\bar{\pi}_{0})^{k_0} (\pi_{\dot{0}})^{l_0}$ 
($k_0, l_0 \in \Bbb{Z}$) in the limit $|\pi_{\dot{0}}| \rightarrow \infty$,  
and as 
$\tilde{f}_{0}^{+}(\bar{\pi}_{0}, \pi_{\dot{0}}) (\bar{\pi}_{1})^{k_1} (\pi_{\dot{1}})^{l_1}$ 
($k_1, l_1 \in \Bbb{Z}$) in the limit $|\pi_{\dot{1}}| \rightarrow \infty$.  
Here, $\tilde{f}_{0}^{+}$ and $\tilde{f}_{1}^{+}$ are complex functions determined from $\tilde{f}^{+}$.  
The function $\tilde{f}^{+}$ is also assumed to be able to include extra constant spinors. 
We consider the positive-frequency spinor wave function defined by 
\begin{align}
\varPsi^{\:\! +}_{\alpha_1 \ldots \alpha_m \dot{\alpha}_1 \ldots \dot{\alpha}_n}(x) 
& :=\frac{(-1)^m}{(2\pi i)^4} \int_{\Bbb{C}^2} \tilde{f}^{+}(\bar{\pi}, \pi) \:\! 
\varPhi_{\alpha_1 \ldots \alpha_m \dot{\alpha}_1 \ldots \dot{\alpha}_n} 
(x, \bar{\pi}, \pi)\:\!  d^2 \bar{\pi} \wedge d^2 \pi 
\notag
\\
&\; =\frac{(-1)^m}{(2\pi i)^4} \int_{\Bbb{C}^2} 
\bar{\pi}_{\alpha_1} \cdots \bar{\pi}_{\alpha_m} 
\pi_{\dot{\alpha}_1} \cdots \pi_{\dot{\alpha}_n} 
\tilde{f}^{+}(\bar{\pi}, \pi) 
\exp \! \Big({-ix^{\beta\dot{\beta}} \bar{\pi}_{\beta} \pi_{\dot{\beta}}} \Big) 
\notag
\\
& \:\:\quad \times d^2 \bar{\pi} \wedge d^2 \pi \,, 
\label{5.1}
\end{align}
where $d^2 \bar{\pi}:=d\bar{\pi}_{0} \wedge d\bar{\pi}_{1}$ and 
$d^2 \pi:=d\pi_{\dot{0}} \wedge d\pi_{\dot{1}}$. 
This function is just a linear combination of 
$\varPhi_{\alpha_1 \ldots \alpha_m \dot{\alpha}_1 \ldots \dot{\alpha}_n}$ with 
the coefficient function $\tilde{f}^{+}$. 
The integral in Eq. (\ref{5.1}) is, however, not well-defined when the absolute value 
of the integrand increases or sufficiently slowly decreases in the asymptotic region 
specified by $|\pi_{\dot{0}}|{}^{2} +|\pi_{\dot{1}}|{}^{2}\rightarrow \infty$.  
(We assume that the integrand behaves well at the origin of $\Bbb{C}^2$.)  
To make this integral well-defined, we now replace $x^{\alpha\dot{\alpha}}$ 
by $z^{\alpha\dot{\alpha}}=x^{\alpha\dot{\alpha}}-iy^{\alpha\dot{\alpha}}$ so that 
the integrand can include the multiplicative exponential factor 
$\exp\!\big({-y^{\beta\dot{\beta}} \bar{\pi}_{\beta} \pi_{\dot{\beta}}} \big)$. 
The exponent $y^{\beta\dot{\beta}} \bar{\pi}_{\beta} \pi_{\dot{\beta}}$ can be written as 
\begin{align}
y^{\beta\dot{\beta}} \bar{\pi}_{\beta} \pi_{\dot{\beta}}
=\frac{1}{\sqrt{2}} \big(y^{0} +|\vec{y}\;\!| \big) |\varpi_{\dot{0}}|^{2} 
+\frac{1}{\sqrt{2}} \big(y^{0} -|\vec{y}\;\!| \big) |\varpi_{\dot{1}}|^{2} 
\label{5.2}
\end{align}
in terms of the real variables $y^{\mu}$ $(\mu=0, 1, 2, 3)$ and the spinor 
$\varpi_{\dot{\alpha}}:=U_{\dot{\alpha}}{}^{\dot{\beta}}(y) \pi_{\dot{\beta}}$.\footnote{~The 
bispinor notation $z^{\alpha \dot{\alpha}}$ and the 4-vector notation $z^{\mu}$  
are related by 
\begin{align}
\begin{pmatrix}
\, z^{0\dot{0}} \:& z^{0\dot{1}} \,\, \\
\, z^{1\dot{0}} \:& z^{1\dot{1}} \,\,
\end{pmatrix}
= \dfrac{1}{\sqrt{2}} \!
\begin{pmatrix}
\, z^{0} + z^{3} \:& z^{1} + i z^{2} \,\, \\
\, z^{1} - i z^{2} \:& z^{0} - z^{3} \;
\end{pmatrix}.
\notag
\end{align}
Note that $z^{\alpha \dot{\alpha}}$ is Hermitian if and only if $z^{\mu}$ is real. 
Since $x^{\alpha \dot{\alpha}}$ and  $y^{\alpha \dot{\alpha}}$ are Hermitian, $x^{\mu}$ and $y^{\mu}$ are real. }
Here, $|\vec{y}\;\!|:=\sqrt{(y^1)^2 +(y^2)^2 +(y^3)^2}\:$ and 
\begin{align}
U(y):=\frac{1}{\sqrt{2|\vec{y}\;\!|\big( y^{3} +|\vec{y}\;\!|\big)}} 
\begin{pmatrix}
\: y^{3} +|\vec{y}\;\!| \;&\; y^{1} + i y^{2} \,\; \\
\: y^{1} - i y^{2} \;&\;  -y^{3} -|\vec{y}\;\!| \,\; 
\end{pmatrix}.
\label{5.3}
\end{align}
This matrix is both unitary and Hermitian.  
From Eq. (\ref{5.2}), we see that $y^{\beta\dot{\beta}} \bar{\pi}_{\beta} \pi_{\dot{\beta}}$ 
is positive definite if and only if 
$y_{\mu} y^{\mu}\equiv (y^0)^2 -|\vec{y}\;\!|^2 >0$ and $y^{0}>0$. 
These two conditions for $y^{\mu}$ together define a region 
called the forward (or future) tube \cite{PM,PR,HT,Tak}:    
\begin{align}
\mathbb{C}\mathbf{M}^{+} := \big\{\:\! (z^{\mu}) \in \mathbb{C}\mathbf{M}^{\sharp} \, \big|\, 
z^{\mu}= x^{\mu} - i y^{\mu}, \, y_{\mu} y^{\mu} > 0\:\!, \, y^{0} > 0  \:\! \big\} \,.   
\label{5.4}
\end{align}
Here, $\mathbb{C}\mathbf{M}^{\sharp}$ denotes the conformal compactification 
of complexified Minkowski space $\mathbb{C}\mathbf{M}$.    
(It becomes possible to involve $(\pi_{\dot{\alpha}})=0$ by taking $\mathbb{C}\mathbf{M}^{\sharp}$  
instead of $\mathbb{C}\mathbf{M}$.)  
Since $y^{\beta\dot{\beta}} \bar{\pi}_{\beta} \pi_{\dot{\beta}}>0$ is valid in $\mathbb{C}\mathbf{M}^{+}$,   
the integral in  
\begin{align}
\varPsi^{\:\!+}_{\alpha_1 \ldots \alpha_m \dot{\alpha}_1 \ldots \dot{\alpha}_n}(z) 
& =\frac{(-1)^{m}}{(2\pi i)^4} \int_{\Bbb{C}^2} 
\bar{\pi}_{\alpha_1} \cdots \bar{\pi}_{\alpha_m} 
\pi_{\dot{\alpha}_1} \cdots \pi_{\dot{\alpha}_n} 
\tilde{f}^{+}(\bar{\pi}, \pi) 
\exp \! \Big({-iz^{\beta\dot{\beta}} \bar{\pi}_{\beta} \pi_{\dot{\beta}}} \Big) 
\notag
\\
& \:\quad \times
d^2 \bar{\pi} \wedge d^2 \pi 
\label{5.5}
\end{align}
is well-defined for $(z^{\mu}) \in \mathbb{C}\mathbf{M}^{+}$. 
Hence, it follows that the positive-frequency spinor wave function 
is properly defined on $\mathbb{C}\mathbf{M}^{+}$.  
In this function, $\exp\!\big({-y^{\beta\dot{\beta}} \bar{\pi}_{\beta} \pi_{\dot{\beta}}} \big)$ plays the role of 
a damping factor. 
The corresponding spinor wave function on $\mathbf{M}$ is given by  
\begin{align}
\varPsi^{\:\! +}_{\alpha_1 \ldots \alpha_m \dot{\alpha}_1 \ldots \dot{\alpha}_n}(x)
:=\lim_{\;\! y^0 \downarrow\:\! 0} 
\varPsi^{\:\! +}_{\alpha_1 \ldots \alpha_m \dot{\alpha}_1 \ldots \dot{\alpha}_n}(z) \,.
\label{5.6}
\end{align}

Using $\bar{\pi}_{\beta} \bar{\pi}^{\beta}=\pi_{\dot{\beta}} \pi^{\dot{\beta}}=0$, we can readily prove that 
$\varPsi^{\:\!+}_{\alpha_1 \ldots \alpha_m \dot{\alpha}_1 \ldots \dot{\alpha}_n}(z)$ satisfies 
the generalized Weyl equations 
\begin{subequations}
\label{5.7}
\begin{align}
\frac{\partial}{\partial z_{\beta\dot{\beta}}} 
\:\! \varPsi_{\beta \alpha_2 \ldots \alpha_m \dot{\alpha}_1 \ldots \dot{\alpha}_n}(z) =0 \,, 
\label{5.7a}
\\
\frac{\partial}{\partial z_{\beta\dot{\beta}}} 
\:\! \varPsi_{\alpha_1 \ldots \alpha_m \dot{\beta} \dot{\alpha}_2 \ldots \dot{\alpha}_n}(z) =0 \,. 
\label{5.7b}
\end{align}
\end{subequations}
We thus see that $\varPsi^{\:\!+}_{\alpha_1 \ldots \alpha_m \dot{\alpha}_1 \ldots \dot{\alpha}_n}(z)$ is a solution of 
the field equations for a free massless spinor field of rank $(m+n)$. 
The function 
$\varPhi_{\alpha_1 \ldots \alpha_m \dot{\alpha}_1 \ldots \dot{\alpha}_n} 
(z, \bar{\pi}, \pi)$ is a particular solution of Eqs. (\ref{5.7a}) and (\ref{5.7b}).

Now, let us consider the Fourier-Laplace transform of $\tilde{f}^{+}(\bar{\pi}, \pi)$ 
with respect to $\bar{\pi}_{\alpha}\:\!$: 
\begin{align}
f^{+}(\omega, \pi) :=\frac{1}{(2\pi i)^{2}} \oint_{\varPi^{+}} 
\tilde{f}^{+}(\bar{\pi}, \pi) 
\exp \! \big(-\bar{\pi}_{\alpha} \omega^{\alpha} \big) 
d^2 \bar{\pi} \,. 
\label{5.8}
\end{align}
Here, $\omega^{\alpha}$ is defined by Eq. (\ref{2.5}), 
and the integral is taken over a suitable 2-dimensional contour, $\varPi^{+}$,  
chosen in such a manner that $f^{+}$ becomes a holomorphic function of $\omega^{\alpha}$ and 
$\pi_{\dot{\alpha}}$. (The Fourier-Laplace transform (\ref{5.8}) is consistent with 
the representation 
$\Hat{\Bar{\pi}}_{\alpha}=-{\partial}/{\partial \omega^{\alpha}}$.) 
Since the pair of $\omega^{\alpha}$ and $\pi_{\dot{\alpha}}$ is precisely the twistor 
$Z^{A}=(\omega^{\alpha}, \pi_{\dot{\alpha}})$, the function $f^{+}$ is regarded as 
a holomorphic function on (non-projective) twistor space $\mathbf{T}$, 
the 4-dimensional complex space coordinatized by $(\omega^{\alpha}, \pi_{\dot{\alpha}})$, 
and can be expressed as $f^{+}(Z)$. 
From the first equality of  
\begin{align}
y^{\beta\dot{\beta}} \bar{\pi}_{\beta} \pi_{\dot{\beta}}=\Re (\bar{\pi}_{\alpha} \omega^{\alpha}) 
=\frac{1}{2} \big(\bar{\pi}_{\alpha} \omega^{\alpha} + \bar{\omega}^{\dot{\alpha}} \pi_{\dot{\alpha}} \big) \,,  
\label{5.9}
\end{align}
it is clear that $f^{+}$ is well-defined on the condition $y^{\beta\dot{\beta}} \bar{\pi}_{\beta} \pi_{\dot{\beta}}>0$.  
In other words, $f^{+}$ is actually well-defined on the upper half of twistor space 
\begin{align}
\mathbf{T}^{+}:= \big\{\:\! (\omega^{\alpha}, \pi_{\dot{\alpha}}) \in \mathbf{T} \, \big|\: 
\bar{\pi}_{\alpha} \omega^{\alpha} + \bar{\omega}^{\dot{\alpha}} \pi_{\dot{\alpha}} >0 \:\! \big\} \,. 
\label{5.10}
\end{align}
This is the region of $\mathbf{T}$ corresponding to $\mathbb{C}\mathbf{M}^{+}$. 
More precisely, an arbitrary point in $\mathbf{T}^{+}$ corresponds to a complex null plane, 
called an $\alpha$-plane, lying entirely in $\mathbb{C}\mathbf{M}^{+}$;  
conversely, an arbitrary point in $\mathbb{C}\mathbf{M}^{+}$ corresponds to 
a 2-dimensional subspace of $\mathbf{T}$ lying entirely in $\mathbf{T}^{+}$.

Noting that $\frac{\partial}{\partial \omega^{\alpha}} 
\exp \! \big(-\bar{\pi}_{\beta} \omega^{\beta} \big) 
=-\bar{\pi}_{\alpha} \exp \! \big(-\bar{\pi}_{\beta} \omega^{\beta} \big)$, 
we can write Eq. (\ref{5.5}) in terms of $f^{+}(\omega, \pi)$ as 
\begin{align}
\varPsi^{\:\!+}_{\alpha_1 \ldots \alpha_m \dot{\alpha}_1 \ldots \dot{\alpha}_n}(z) 
=\frac{1}{(2\pi i)^2} \oint_{\varSigma^{+}} 
\pi_{\dot{\alpha}_1} \cdots \pi_{\dot{\alpha}_n} 
\frac{\partial}{\partial \omega^{\alpha_1}} \cdots \frac{\partial}{\partial \omega^{\alpha_m}}
f^{+}(\omega, \pi) d^2 \pi \,, 
\label{5.11}
\end{align}
where $(z^{\mu}) \in \mathbb{C}\mathbf{M}^{+}$, 
$(\omega^{\alpha}, \pi_{\dot{\alpha}}) \in \mathbf{T}^{+}$, and 
$\varSigma^{+}$ is another 2-dimensional contour. 
Equation (\ref{5.11}) is precisely 
the one known as the non-projective form of the Penrose transform 
with both dotted and undotted spinor indices \cite{PR}.   
The exterior derivative of the integrand  
including $d^2 \pi$ vanishes with $z^{\mu}$ held constant:  
\begin{align}
d \bigg( \pi_{\dot{\alpha}_1} \cdots \pi_{\dot{\alpha}_n} 
\frac{\partial}{\partial \omega^{\alpha_1}} \cdots \frac{\partial}{\partial \omega^{\alpha_m}}
f^{+}(\omega, \pi) d^2 \pi \bigg)=0 \,.
\label{5.11.1}
\end{align} 
Therefore it can be proven by using Poincar\'{e}'s lemma and Stokes' theorem that  
the integral itself remains invariant under the deformations of $\varSigma^{+}$ 
that are carried out continuously in the domain of the integrand. 
Suppose now that $f^{+}$ is homogeneous of degree $r$, that is,   
$f^{+}(c\:\!\omega, c\:\!\pi)=c^{r} f^{+}(\omega, \pi)\:\!$ ($c \in \Bbb{C}$). 
Then, under the replacement of $\pi_{\dot{\alpha}}$ by $c\:\!\pi_{\dot{\alpha}}$,  
the integral changes into that multiplied by $c^{n-m+r+2}$ by virtue of the deformation 
invariance of the integral. 
However, this replacement cannot change the integral actually, 
because the $\pi_{\dot{\alpha}}$ are merely variables of integration. 
Hence, it follows that the integral vanishes if $r \neq m-n-2\:\!$;  
only in the case $r=m-n-2$, the integral may remain non-vanishing. 
In this case, the integrand including $d^2 \pi$ can be expressed as the exterior product of 
$d\pi_{\dot{0}}/\pi_{\dot{0}}$ and a 1-form consisting of $\zeta:=\pi_{\dot{1}}/\pi_{\dot{0}}$. 
(Here, $\pi_{\dot{0}}$ and $\zeta$ are treated as independent variables.) 
After carrying out the contour integration over $\pi_{\dot{0}}$ along a topological circle 
that surrounds $\pi_{\dot{0}}=0$, Eq. (\ref{5.11}) reduces to 
\begin{align}
\varPsi^{\:\!+}_{\alpha_1 \ldots \alpha_m \dot{\alpha}_1 \ldots \dot{\alpha}_n}(z) 
=\frac{1}{2\pi i} \oint_{\varGamma^{+}} 
\pi_{\dot{\alpha}_1} \cdots \pi_{\dot{\alpha}_n} 
\frac{\partial}{\partial \omega^{\alpha_1}} \cdots \frac{\partial}{\partial \omega^{\alpha_m}}
f^{+}(\omega, \pi) \pi_{\dot{\beta}} d\pi^{\dot{\beta}} \:\! , 
\label{5.12}
\end{align}
where $\varGamma^{+}$ denotes a 1-dimensional closed contour on 
the complex projective line $\Bbb{C}\mathbf{P}^{1}$ coordinatized by $\zeta$ or $\zeta^{-1}$. 
Equation (\ref{5.12}) is known as the projective form of the Penrose transform 
\cite{PM, PR, HT, Tak, Pen}. 
It is easy to show that $f^{+}$ satisfies an analog of the helicity eigenvalue equation: 
\begin{align}
\left( -\omega^{\alpha} \frac{\partial}{\partial {\omega}_{\alpha}}
-\pi_{\dot{\alpha}} \frac{\partial}{\partial \pi_{\dot{\alpha}}} -2+2s \right) 
\! f^{+}(\omega, \pi) &=0 \,. 
\label{5.13}
\end{align}
This looks like the helicity eigenvalue equation treated in twistor theory, 
but the sign of $s$ is opposite to that of the usual one.\footnote{~The 
helicity eigenvalue equation treated in twistor theory is given by \cite{PM,PR} 
\begin{align*}
\left( -\omega^{\alpha} \frac{\partial}{\partial {\omega}_{\alpha}}
-\pi_{\dot{\alpha}} \frac{\partial}{\partial \pi_{\dot{\alpha}}} -2-2s \right) 
\! f (\omega, \pi) &=0 \,. 
\end{align*}
This can be derived from the action (\ref{2.4}) through the twistor quantization procedure 
and can be shown to be equivalent to Eq. (\ref{4.8c}). 
}
Hence, it turns out that $\varPsi^{\:\!+}_{\dot{\alpha}_1 \ldots \dot{\alpha}_n}$ describes a negative 
helicity field, while $\varPsi^{\:\!+}_{\alpha_1 \ldots \alpha_m}$ describes a positive helicity field. 
This result can also be seen from Eq. (\ref{4.10}).

\subsection{Negative-frequency wave function} 

A (well-defined) negative-frequency spinor wave function can be obtained immediately by taking 
the complex conjugate of 
$\varPsi^{\:\! +}_{\alpha_1 \ldots \alpha_m \dot{\alpha}_1 \ldots \dot{\alpha}_n}(z)$. 
The wave function obtained in this manner is, however, a function of $\overline{z^{\mu}}$ and 
hence is anti-holomorphic. 
In the following, we construct a {\em holomorphic} negative-frequency spinor wave function.

Let $\tilde{f}^{-}(\bar{\pi}, \pi)$ be a complex function similar to $\tilde{f}^{+}(\bar{\pi}, \pi)$. 
The negative-frequency counterpart of  
$\varPsi^{\:\!+}_{\alpha_1 \ldots \alpha_m \dot{\alpha}_1 \ldots \dot{\alpha}_n}(x)$ in Eq. (\ref{5.1}) is 
defined by
\begin{align}
\varPsi^{\:\! -}_{\alpha_1 \ldots \alpha_m \dot{\alpha}_1 \ldots \dot{\alpha}_n}(x) 
& :=\frac{1}{(2\pi i)^4} \int_{\Bbb{C}^2} \tilde{f}^{-}(\bar{\pi}, \pi) \:\! 
\bar{\varPhi}_{\alpha_1 \ldots \alpha_m \dot{\alpha}_1 \ldots \dot{\alpha}_n} 
(x, \bar{\pi}, \pi)\:\!  d^2 \bar{\pi} \wedge d^2 \pi 
\notag
\\
&\; =\frac{1}{(2\pi i)^4} \int_{\Bbb{C}^2} 
\bar{\pi}_{\alpha_1} \cdots \bar{\pi}_{\alpha_m} 
\pi_{\dot{\alpha}_1} \cdots \pi_{\dot{\alpha}_n} 
\tilde{f}^{-}(\bar{\pi}, \pi) 
\exp \! \Big({ix^{\beta\dot{\beta}} \bar{\pi}_{\beta} \pi_{\dot{\beta}}} \Big) 
\notag
\\
& \:\:\quad \times d^2 \bar{\pi} \wedge d^2 \pi \,, 
\label{5.14}
\end{align}
where 
\begin{align}
\bar{\varPhi}_{\alpha_1 \ldots \alpha_m \dot{\alpha}_1 \ldots \dot{\alpha}_n} 
(x, \bar{\pi}, \pi) :=
\overline{{\varPhi}_{\alpha_1 \ldots \alpha_n \dot{\alpha}_1 \ldots \dot{\alpha}_m} 
(x, \bar{\pi}, \pi)} \,. 
\label{5.15}
\end{align}
The function $\bar{\varPhi}_{\alpha_1 \ldots \alpha_m \dot{\alpha}_1 \ldots \dot{\alpha}_n}$ obeys 
the complex conjugates of Eqs. (\ref{4.8a})--(\ref{4.8c}), and its corresponding value of $s$ is 
determined to be 
\begin{align}
s=-\frac{1}{2}(m-n) \,, \quad\; m,n=0,1,2,\ldots. 
\label{5.16}
\end{align}
Note that this is different from Eq. (\ref{4.10}) only in the sign. 
The integral in Eq. (\ref{5.14}) itself is not well-defined in general, and 
we therefore replace $x^{\alpha\dot{\alpha}}$ 
with $z^{\alpha\dot{\alpha}}=x^{\alpha\dot{\alpha}}-iy^{\alpha\dot{\alpha}}$ by    
following the case of the positive-frequency spinor wave function. 
Owing to the replacement, the integrand is modified so as to include the damping factor 
$\exp\!\big({y^{\beta\dot{\beta}} \bar{\pi}_{\beta} \pi_{\dot{\beta}}} \big)$  
valid on the simultaneous conditions $y_{\mu} y^{\mu}>0$ and $y^{0}<0$. 
(Recall here Eq. (\ref{5.2}).) 
These conditions together define a region called  
the backward (or past) tube \cite{PM,PR,HT,Tak}:    
\begin{align}
\mathbb{C}\mathbf{M}^{-} := \big\{\:\! (z^{\mu}) \in \mathbb{C}\mathbf{M}^{\sharp} \, \big|\, 
z^{\mu}= x^{\mu} - i y^{\mu}, \, y_{\mu} y^{\mu} > 0\:\!, \, y^{0} <0  \:\! \big\} \,.   
\label{5.17}
\end{align}
Since $y^{\beta\dot{\beta}} \bar{\pi}_{\beta} \pi_{\dot{\beta}}<0$ is fulfilled in $\mathbb{C}\mathbf{M}^{-}$,   
the integral in  
\begin{align}
\varPsi^{\:\!-}_{\alpha_1 \ldots \alpha_m \dot{\alpha}_1 \ldots \dot{\alpha}_n}(z) 
&=\frac{1}{(2\pi i)^4} \int_{\Bbb{C}^2} 
\bar{\pi}_{\alpha_1} \cdots \bar{\pi}_{\alpha_m} 
\pi_{\dot{\alpha}_1} \cdots \pi_{\dot{\alpha}_n} 
\tilde{f}^{-}(\bar{\pi}, \pi) 
\exp \! \Big({iz^{\beta\dot{\beta}} \bar{\pi}_{\beta} \pi_{\dot{\beta}}} \Big)
\notag
\\
& \:\quad \times
d^2 \bar{\pi} \wedge d^2 \pi 
\label{5.18}
\end{align}
is well-defined for $(z^{\mu}) \in \mathbb{C}\mathbf{M}^{-}$. 
It thus follows that the holomorphic negative-frequency spinor wave function 
is properly defined on $\mathbb{C}\mathbf{M}^{-}$. 
The corresponding spinor wave function on $\mathbf{M}$ is given by  
\begin{align}
\varPsi^{\:\! -}_{\alpha_1 \ldots \alpha_m \dot{\alpha}_1 \ldots \dot{\alpha}_n}(x)
:=\lim_{\;\! y^0 \uparrow\:\! 0} 
\varPsi^{\:\! -}_{\alpha_1 \ldots \alpha_m \dot{\alpha}_1 \ldots \dot{\alpha}_n}(z) \,.
\label{5.19}
\end{align}
We can easily prove that  
$\varPsi^{\:\!-}_{\alpha_1 \ldots \alpha_m \dot{\alpha}_1 \ldots \dot{\alpha}_n}(z)$ 
satisfies the generalized Weyl equations (\ref{5.7a}) and (\ref{5.7b}).

Next, we consider the Fourier-Laplace transform of $\tilde{f}^{-}(\bar{\pi}, \pi)$ 
with respect to $\bar{\pi}_{\alpha}\:\!$: 
\begin{align}
f^{-}(\omega, \pi) :=\frac{1}{(2\pi i)^{2}} \oint_{\varPi^{-}} 
\tilde{f}^{-}(\bar{\pi}, \pi) 
\exp \! \big(\bar{\pi}_{\alpha} \omega^{\alpha} \big) 
d^2 \bar{\pi} \,. 
\label{5.20}
\end{align}
Here, the integral is taken over a suitable 2-dimensional contour, $\varPi^{-}$,  
chosen in such a manner that $f^{-}$ becomes a holomorphic function of $\omega^{\alpha}$ and 
$\pi_{\dot{\alpha}}$. 
(The Fourier-Laplace transform (\ref{5.20}) is consistent with the {\em conjugate} representation 
$\Hat{\Bar{\pi}}_{\alpha}={\partial}/{\partial \omega^{\alpha}}$.) 
It is clear from Eq. (\ref{5.9}) that $f^{-}$ is well-defined on the lower half of twistor space 
\begin{align}
\mathbf{T}^{-}:= \big\{\:\! (\omega^{\alpha}, \pi_{\dot{\alpha}}) \in \mathbf{T} \, \big|\: 
\bar{\pi}_{\alpha} \omega^{\alpha} + \bar{\omega}^{\dot{\alpha}} \pi_{\dot{\alpha}} <0 \:\! \big\} \,. 
\label{5.21}
\end{align}
This is the region of $\mathbf{T}$ corresponding to $\mathbb{C}\mathbf{M}^{-}$;  
a correspondence similar to that between $\mathbf{T}^{+}$ and $\mathbb{C}\mathbf{M}^{+}$ 
is established between $\mathbf{T}^{-}$ and $\mathbb{C}\mathbf{M}^{-}$.

We can write Eq. (\ref{5.18}) in terms of $f^{-}(\omega, \pi)$ as 
\begin{align}
\varPsi^{\:\!-}_{\alpha_1 \ldots \alpha_m \dot{\alpha}_1 \ldots \dot{\alpha}_n}(z) 
=\frac{1}{(2\pi i)^2} \oint_{\varSigma^{-}} 
\pi_{\dot{\alpha}_1} \cdots \pi_{\dot{\alpha}_n} 
\frac{\partial}{\partial \omega^{\alpha_1}} \cdots \frac{\partial}{\partial \omega^{\alpha_m}}
f^{-}(\omega, \pi) d^2 \pi \,, 
\label{5.22}
\end{align}
where $(z^{\mu}) \in \mathbb{C}\mathbf{M}^{-}$, 
$(\omega^{\alpha}, \pi_{\dot{\alpha}}) \in \mathbf{T}^{-}$, and 
$\varSigma^{-}$ is another 2-dimensional contour. 
Suppose now that $f^{-}$ is homogeneous of degree $r^{\prime}$.  
Then, if $r^{\prime}\neq m-n-2$, the integral vanishes; 
if $r^{\prime}=m-n-2$, the integral may remain non-vanishing 
and can be written as 
\begin{align}
\varPsi^{\:\!-}_{\alpha_1 \ldots \alpha_m \dot{\alpha}_1 \ldots \dot{\alpha}_n}(z) 
=\frac{1}{2\pi i} \oint_{\varGamma^{-}} 
\pi_{\dot{\alpha}_1} \cdots \pi_{\dot{\alpha}_n} 
\frac{\partial}{\partial \omega^{\alpha_1}} \cdots \frac{\partial}{\partial \omega^{\alpha_m}}
f^{-}(\omega, \pi) \pi_{\dot{\beta}} d\pi^{\dot{\beta}} \:\! , 
\label{5.23}
\end{align}
where $\varGamma^{-}$ denotes a 1-dimensional closed contour on $\Bbb{C}\mathbf{P}^{1}$. 
In this way, we obtain the negative-frequency wave function 
$\varPsi^{\:\!-}_{\alpha_1 \ldots \alpha_m \dot{\alpha}_1 \ldots \dot{\alpha}_n}(z)$ 
written in the form of a Penrose transform. 
We can show that $f^{-}$ satisfies an analog of the helicity eigenvalue equation in the conjugate representation: 
\begin{align}
\left( \omega^{\alpha} \frac{\partial}{\partial {\omega}_{\alpha}}
+\pi_{\dot{\alpha}} \frac{\partial}{\partial \pi_{\dot{\alpha}}} +2+2s \right) 
\!\!\: f^{-}(\omega, \pi) &=0 \,, 
\label{5.24}
\end{align}
where $s$ is given in Eq. (\ref{5.16}).\footnote{~The helicity eigenvalue equation  
in the conjugate representation is found to be  
\begin{align*}
\left( \omega^{\alpha} \frac{\partial}{\partial {\omega}_{\alpha}}
+\pi_{\dot{\alpha}} \frac{\partial}{\partial \pi_{\dot{\alpha}}} +2-2s \right) 
\!\!\: g (\omega, \pi) &=0 \,, 
\end{align*}
which is different from Eq. (\ref{5.24}) only in the sign of $s$. 
This equation can be derived from the action (\ref{2.4}) through the twistor quantization procedure 
and can be shown to be equivalent to the complex conjugate of Eq. (\ref{4.8c}), namely 
\begin{align*}
\left( \pi_{\dot{\alpha}} \frac{\partial}{\partial \pi_{\dot{\alpha}}}
-\bar{\pi}_{\alpha} \frac{\partial}{\partial \bar{\pi}_{\alpha}}
- 2s \right) 
\!\!\: \bar{\varPhi}(\bar{\pi}, \pi) &=0 \,. 
\end{align*}
}
As can be seen from Eq. (\ref{5.16}), $\varPsi^{\:\!-}_{\dot{\alpha}_1 \ldots \dot{\alpha}_n}$ describes a positive  
helicity field, while $\varPsi^{\:\!-}_{\alpha_1 \ldots \alpha_m}$ describes a negative helicity field.

\section{Exponential generating function for the spinor wave functions} 

In this section, we consider the exponential generating function for the spinor wave functions 
defined in Sec. 5. From the generating function, 
we derive a novel representation for each of the spinor wave functions.

From Eqs. (\ref{5.5}) and (\ref{5.18}),  it is easily seen that 
\begin{align}
-i \frac{\partial}{\partial z^{\beta \dot{\beta}}} 
\:\! \varPsi^{\:\!\pm}_{\alpha_1 \ldots \alpha_m \dot{\alpha}_1 \ldots \dot{\alpha}_n}(z) 
=\varPsi^{\:\!\pm}_{\beta \alpha_1 \ldots \alpha_m \dot{\beta} \dot{\alpha}_1 \ldots \dot{\alpha}_n}(z) \,.
\label{6.1}
\end{align}
Now we define the exponential generating function, $\boldsymbol{\varPsi}^{\:\!\pm}$, 
for the spinor wave functions 
$\varPsi^{\:\!\pm}_{\alpha_1 \ldots \alpha_m \dot{\alpha}_1 \ldots \dot{\alpha}_n}(z)$:  
\begin{align}
\boldsymbol{\varPsi}^{\:\!\pm}(z, \iota, \kappa)
:=\sum_{m=0}^{\infty} \sum_{n=0}^{\infty} \frac{1}{m! \:\! n!}  
\:\! \varPsi^{\:\!\pm}_{\alpha_1 \ldots \alpha_m \dot{\alpha}_1 \ldots \dot{\alpha}_n}(z) 
\iota^{\alpha_1} \cdots \iota^{\alpha_m} 
\kappa^{\dot{\alpha}_1} \cdots \kappa^{\dot{\alpha}_n} , 
\label{6.2}
\end{align}
where $\iota^{\alpha}$ and $\kappa^{\dot{\alpha}}$ are arbitrary undotted and dotted spinors, respectively. 
The functions $\varPsi^{\:\!\pm}_{\alpha_1 \ldots \alpha_m \dot{\alpha}_1 \ldots \dot{\alpha}_n}(z)$ 
can be treated as expansion coefficients in the Maclaurin series expansion of $\boldsymbol{\varPsi}^{\:\!\pm}$ 
with respect to $\iota^{\alpha}$ and $\kappa^{\dot{\alpha}}$. 
Using Eq. (\ref{6.1}), we can show that $\boldsymbol{\varPsi}^{\:\!\pm}$ satisfies the fundamental equation 
\begin{align}
\left( -i\frac{\partial}{\partial z^{\alpha\dot{\alpha}}}
-\frac{\partial^2}{\partial \iota^{\alpha} \partial \kappa^{\dot{\alpha}}}
\right) \:\!\! \boldsymbol{\varPsi}^{\:\!\pm}(z, \iota, \kappa)=0 \,. 
\label{6.3}
\end{align}
This is precisely the complexification of the so-called {\em unfolded equation} \cite{FedIva,Vas} 
\begin{align}
\left( -i\frac{\partial}{\partial x^{\alpha\dot{\alpha}}}
-\frac{\partial^2}{\partial \psi^{\alpha} \partial \bar{\psi}^{\dot{\alpha}}}
\right) \:\!\! \tilde{\varPhi} \big(x, \psi, \bar{\psi} \big)=0 \,, 
\label{6.4}
\end{align}
which can be obtained in the present formulation by taking the inner product between 
Eq. (\ref{4.3a}) and the bra-vector   
\begin{align}
\langle x, a, \psi, \bar{\psi} |
:=\langle \tilde{0} | \exp \!\left( ix^{\alpha\dot{\alpha}} \hat{P}_{\alpha \dot{\alpha}}^{(x)}
+ia \hat{P}^{(a)} 
-\psi^{\alpha} \Hat{\bar{\pi}}_{\alpha} +\Bar{\psi}{}^{\dot{\alpha}} \Hat{\pi}_{\dot{\alpha}}  \right) . 
\label{6.5}
\end{align}
Here, $\langle \tilde{0} |$ is a reference bra-vector specified by 
$\langle \tilde{0} |\:\!\hat{x}{}^{\alpha \dot{\alpha}}=\langle \tilde{0} |\hat{a} 
=\langle \tilde{0} | \hat{\psi}{}^{\alpha} =\langle \tilde{0} |\Hat{\Bar{\psi}}{}^{\dot{\alpha}} =0$.  
The function $\tilde{\varPhi}$ is defined by 
$\tilde{\varPhi}(x, a, \psi, \bar{\psi}):=\langle x, a, \psi, \bar{\psi} |\varPhi \rangle$ 
and is described as $\tilde{\varPhi}(x, \psi, \bar{\psi})$ after taking into account Eq. (\ref{4.3b}).

Substituting Eqs. (\ref{5.5}) and (\ref{5.18}) into (\ref{6.2}), we have 
\begin{align}
\boldsymbol{\varPsi}^{\:\!\pm}(z, \iota, \kappa)
=\frac{1}{(2\pi i)^4} \int_{\Bbb{C}^2} 
\tilde{f}^{\pm}(\bar{\pi}, \pi) 
\exp \! \big(\mp iz^{\alpha\dot{\alpha}} \bar{\pi}_{\alpha} \pi_{\dot{\alpha}} 
\mp \bar{\pi}_{\alpha} \iota^{\alpha} +\pi_{\dot{\alpha}} \kappa^{\dot{\alpha}} \big)
d^2 \bar{\pi} \wedge d^2 \pi \,. 
\label{6.6}
\end{align}
With this expression, it is clear that $\boldsymbol{\varPsi}^{\:\!+}$ and $\boldsymbol{\varPsi}^{\:\!-}$ 
are well-defined on $\mathbb{C}\mathbf{M}^{+}$ and $\mathbb{C}\mathbf{M}^{-}$, respectively, 
owing to the fact that the integrals converge in their corresponding tube domains. 
Substitution of Eqs. (\ref{5.12}) and (\ref{5.23}) into Eq. (\ref{6.2}) yields  
\begin{align}
\boldsymbol{\varPsi}^{\:\!\pm}(z, \iota, \kappa)
=\frac{1}{2\pi i} \oint_{\varGamma^{\pm}} 
\exp \!\bigg( \pi_{\dot{\alpha}} \kappa^{\dot{\alpha}} +\iota^{\alpha} \frac{\partial}{\partial \omega^{\alpha}} \bigg) 
f^{\pm}(\omega, \pi) \pi_{\dot{\beta}} d\pi^{\dot{\beta}} \:\! , 
\label{6.7}
\end{align}
which can be recognized as a collective form of the Penrose transforms found in Sec. 5.

We now note that 
\begin{align}
\tilde{f}^{\pm}(\bar{\pi}, \pi) 
\exp \! \big( \mp \bar{\pi}_{\alpha} \iota^{\alpha} +\pi_{\dot{\alpha}} \kappa^{\dot{\alpha}} \big)
=\tilde{f}^{\pm} \!\!\; \bigg(\! \mp  \!\!\:\frac{\partial}{\partial \iota}, \:\! \frac{\partial}{\partial \kappa} \bigg) 
\exp \! \big( \mp \bar{\pi}_{\alpha} \iota^{\alpha} +\pi_{\dot{\alpha}} \kappa^{\dot{\alpha}} \big) \,, 
\label{6.8}
\end{align}
where $\tilde{f}^{\pm} \big(\! \mp \! \partial/\partial \iota, \:\! \partial/\partial \kappa \big)$ may include  
the integration operators $(\partial/\partial \iota^{\alpha})^{-1}:=\int d \iota^{\alpha}$ and   
$(\partial/\partial \kappa^{\dot{\alpha}})^{-1}:=\int d \kappa^{\dot{\alpha}}$, and their higher-order analogs.  
Applying Eq. (\ref{6.8}) to Eq. (\ref{6.6}), we obtain 
\begin{align}
\boldsymbol{\varPsi}^{\:\!\pm}(z, \iota, \kappa)
&=\frac{1}{(2\pi i)^4} \:\! 
\tilde{f}^{\pm} \!\!\; \bigg(\! \mp \!\!\: \frac{\partial}{\partial \iota}, \:\! \frac{\partial}{\partial \kappa} \bigg) 
\int_{\Bbb{C}^2} 
\exp \! \big(\mp iz^{\alpha\dot{\alpha}} \bar{\pi}_{\alpha} \pi_{\dot{\alpha}} 
\mp \bar{\pi}_{\alpha} \iota^{\alpha} +\pi_{\dot{\alpha}} \kappa^{\dot{\alpha}} \big) 
\notag
\\
& \: \quad \times d^2 \bar{\pi} \wedge d^2 \pi
\notag
\\
&=\frac{1}{(2\pi i)^4} \:\!
\tilde{f}^{\pm} \!\!\; \bigg(\! \mp  \!\!\: \frac{\partial}{\partial \iota}, \:\! \frac{\partial}{\partial \kappa} \bigg) 
\exp \! \big( i z^{-1}_{\dot{\alpha} \alpha} \kappa^{\dot{\alpha}} \iota^{\alpha} \big) 
\int_{\Bbb{C}^2} 
\exp \! \big(\mp iz^{\beta\dot{\beta}} \bar{\pi}_{\beta} \pi_{\dot{\beta}} \big)
\notag
\\
& \: \quad \times d^2 \bar{\pi} \wedge d^2 \pi \,. 
\label{6.9}
\end{align}
Here, $z^{-1}_{\dot{\alpha} \alpha}$ denote the matrix elements such that 
$z^{\alpha\dot{\gamma}} z^{-1}_{\dot{\gamma} \beta}=\delta^{\alpha}{}_{\beta}$ 
and $z^{-1}_{\dot{\alpha} \gamma} z^{\gamma \dot{\beta}} 
=\delta_{\dot{\alpha}}{}^{\dot{\beta}}$. 
Carrying out the integration in (\ref{6.9}) leads to 
\begin{align}
\boldsymbol{\varPsi}^{\:\!\pm}(z, \iota, \kappa)&=\frac{1}{(2\pi)^{2}} 
\det \!\Big(z^{-1}_{\dot{\beta} \beta} \Big) \:\!
\tilde{f}^{\pm} \!\!\; \bigg(\! \mp \!\!\: \frac{\partial}{\partial \iota}, \:\! \frac{\partial}{\partial \kappa} \bigg) 
\exp \! \big( i z^{-1}_{\dot{\alpha} \alpha} \kappa^{\dot{\alpha}} \iota^{\alpha} \big) \,.
\label{6.10}
\end{align}
We can directly verify that $\boldsymbol{\varPsi}^{\:\!\pm}$ in Eq. (\ref{6.10}) fulfills Eq. (\ref{6.3}). 
The spinor wave functions can be derived from Eq. (\ref{6.10}) as 
the coefficients of the Maclaurin series expansion of $\boldsymbol{\varPsi}^{\:\!\pm}$ 
with respect to $\iota^{\alpha}$ and $\kappa^{\dot{\alpha}}\:\!$:  
\begin{align}
\varPsi^{\:\!\pm}_{\alpha_1 \ldots \alpha_m \dot{\alpha}_1 \ldots \dot{\alpha}_n}(z) 
&=\frac{1}{(2\pi)^{2}} \det \!\Big(z^{-1}_{\dot{\beta} \beta} \Big) 
\frac{\partial^{m+n}}{
\partial\iota^{\alpha_1} \cdots \partial\iota^{\alpha_m} 
\partial\kappa^{\dot{\alpha}_1} \cdots \partial\kappa^{\dot{\alpha}_n}} 
\notag
\\
&\quad \;\times
\tilde{f}^{\pm} \!\!\; \bigg(\! \mp \!\!\: \frac{\partial}{\partial \iota}, \:\! \frac{\partial}{\partial \kappa} \bigg) 
\exp \! \big( i z^{-1}_{\dot{\alpha} \alpha} \kappa^{\dot{\alpha}} \iota^{\alpha} \big) 
\bigg|_{\iota^{\alpha}=\kappa^{\dot{\alpha}}=0} \,.
\label{6.11}
\end{align}
In this way, we have obtained a novel representation for each of the spinor wave functions.   
We now write the contravariant vector 
corresponding to $z^{-1}_{\dot{\alpha} \alpha}$ as $(z^{-1})^{\mu}$.\footnote{~The 
bispinor $z_{\dot{\alpha} \beta}  (:=z_{\beta\dot{\alpha}})$ 
is related to the contravariant vector $z^{\mu}$ by 
\begin{align}
\begin{pmatrix}
\, z_{\dot{0} 0} \:& z_{\dot{0} 1} \,\, \\
\, z_{\dot{1} 0} \:& z_{\dot{1} 1} \,\,
\end{pmatrix}
=
\begin{pmatrix}
\, z_{0\dot{0}} \:& z_{1\dot{0}} \,\, \\
\, z_{0\dot{1}} \:& z_{1\dot{1}} \,\,
\end{pmatrix}
= \dfrac{1}{\sqrt{2}} \!
\begin{pmatrix}
\, z^{0} - z^{3} \:& -z^{1} - i z^{2} \,\, \\
\, -z^{1} + i z^{2} \:& z^{0} + z^{3} \;
\end{pmatrix}.
\notag
\end{align}
}
Then it can be shown that $(z^{-1})^{\mu}=2z^{\mu}/(z_{\nu} z^{\nu})$. 
The discrete transformation $z^{\mu} \rightarrow \frac{1}{2}(z^{-1})^{\mu}$ is known as 
the conformal inversion transformation \cite{FraPal}. 
Therefore it turns out that 
$\varPsi^{\:\!\pm}_{\alpha_1 \ldots \alpha_m \dot{\alpha}_1 \ldots \dot{\alpha}_n}$ in Eq. (\ref{6.11}) 
is a function of the conformally inverted spacetime variables 
$\frac{1}{2}(z^{-1})^{\mu}$.

\section{Summary and discussion}

We have studied the canonical formalism and quantization of the model specified by 
the gauged Shirafuji action written in terms of spacetime and spinor variables. 
After a brief review on the gauged Shirafuji action, we investigated the constrained Hamiltonian system 
defined by this action via a Legendre transform 
and systematically classified the constraints into first and second classes. 
In the subsequent quantization procedure, we obtained the plane-wave solution 
$\varPhi_{\alpha_1 \ldots \alpha_m \dot{\alpha}_1 \ldots \dot{\alpha}_n}$ 
by simultaneously solving a set of the physical state conditions based on the first-class constraints. 
We constructed the positive-frequency spinor wave function 
$\varPsi^{\:\! +}_{\alpha_1 \ldots \alpha_m \dot{\alpha}_1 \ldots \dot{\alpha}_n}$ as a linear combination of 
$\varPhi_{\alpha_1 \ldots \alpha_m \dot{\alpha}_1 \ldots \dot{\alpha}_n}$ 
with a coefficient function $\tilde{f}^{+}$ and constructed the negative-frequency counterpart 
$\varPsi^{\:\!-}_{\alpha_1 \ldots \alpha_m \dot{\alpha}_1 \ldots \dot{\alpha}_n}$ 
as a linear combination of $\bar{\varPhi}_{\alpha_1 \ldots \alpha_m \dot{\alpha}_1 \ldots \dot{\alpha}_n}$ 
with a coefficient function $\tilde{f}^{-}$. 
It was shown that 
$\varPsi^{\:\! +}_{\alpha_1 \ldots \alpha_m \dot{\alpha}_1 \ldots \dot{\alpha}_n}$ and 
$\varPsi^{\:\!-}_{\alpha_1 \ldots \alpha_m \dot{\alpha}_1 \ldots \dot{\alpha}_n}$ 
are well-defined on the forward tube $\mathbb{C}\mathbf{M}^{+}$ 
and the backward tube $\mathbb{C}\mathbf{M}^{-}$, respectively, and fulfill the generalized 
Weyl equations (\ref{5.7a}) and (\ref{5.7b}). 
Also, it was demonstrated that the spinor wave functions can be expressed as the Penrose transforms 
of the holomorphic functions $f^{+}$ and $f^{-}$ that are defined as the Fourier-Laplace transforms of 
$\tilde{f}^{+}$ and $\tilde{f}^{-}$, respectively.  
In this way, we have succeeded in finding Penrose transforms via appropriate Fourier-Laplace transforms. 
Furthermore, we constructed the exponential generating function $\boldsymbol{\varPsi}^{\:\!\pm}$ 
for the spinor wave functions 
and derived from it a novel representation, Eq.  (\ref{6.11}), for each of 
the spinor wave functions.  
Then this representation turned out to be a function of the conformally inverted spacetime variables 
$\frac{1}{2}(z^{-1})^{\mu}$.

In this paper, we have optimistically considered 
the existence of the 2-dimensional contour integrals in Eqs. (\ref{5.8}), (\ref{5.11}), (\ref{5.20}) and (\ref{5.22}).  
For making our argument solid,  
it is necessary to prove the existence of these contour integrals using mathematical tools 
developed in higher-dimensional complex analysis. 
In addition, we have not examined function spaces, 
or (pre-)Hilbert spaces, on which Eqs. (\ref{4.6}) and (\ref{4.7}) are supported.  
Since (pre-)Hilbert spaces in twistor quantization have been clarified in Ref. \citen{DN}, 
we expect that the representation defined by Eqs. (\ref{4.6}) and (\ref{4.7}) can be justified 
in these spaces.  
It is also important to see how the Fourier-Laplace transforms in Eqs. (\ref{5.8}) and (\ref{5.20}) 
are realized in the (pre-)Hilbert spaces.

We have treated only free massless particles. 
As a next task, we would like to incorporate interactions with gauge fields lying in spacetime 
into the present twistor model. 
Also, it will be interesting to generalize the gauged Shirafuji action to describe massive spinning particles.  
(An earlier study of this generalization has already been reported in Ref. \citen{Deg}. 
Twistor approaches to describing massive spinning particles have also be presented, for instance,  
in Refs. \citen{BP, Bett, FZ, FFLM}. )

\section*{Acknowledgments}
We are grateful to Shigefumi Naka, Takeshi Nihei and Akitsugu Miwa   
for their useful comments.  
We would like to thank Igor A. Bandos, Sergey Fedoruk and Mikhail S. Plyushchay 
for informing us their related papers.  
One of us (T. Suzuki) thanks Kenji Yamada, Katsuhito Yamaguchi and Haruki Toyoda 
for their encouragement. 
The work of S. Deguchi is supported in part by  
Grant-in-Aid for Fundamental Scientific Research from   
College of Science and Technology, Nihon University.


\end{document}